\documentclass[%
twocolumn,
print,
superscriptaddress,
amsmath,
amssymb,
aps,
rmp,
floatfix,
]{revtex4-2}

\usepackage{graphicx}
\usepackage{dcolumn}
\usepackage{bm}
\usepackage{hyperref}
\usepackage[mathlines]{lineno}

\usepackage{stmaryrd} 
\usepackage{xcolor, soul}
\usepackage{placeins} 
\usepackage{subcaption} 
\usepackage{enumerate} 
\usepackage{yhmath} 
\usepackage{enumitem} 

\begin{document}
\preprint{APS/123-QED}

\title{A minimal model of money creation within secured interbank markets}

\author{Victor Le Coz}
\email{victor.lecoz@gmail.com}
\affiliation{Quant AI Lab, 29 Rue de Choiseul 75002 Paris, France}
\affiliation{Chair of Econophysics and Complex Systems, \'Ecole polytechnique, 91128 Palaiseau Cedex, France}
\affiliation{LadHyX UMR CNRS 7646, \'Ecole polytechnique, 91128 Palaiseau Cedex, France}
\affiliation{Laboratoire de Math\'ematiques et Informatique pour la Complexit\'e et les Syst\`emes, CentraleSupélec, Universit\'e Paris-Saclay, 91192 Gif-sur-Yvette Cedex, France}

\author{Michael Benzaquen}
\email{michael.benzaquen@polytechnique.edu}
\affiliation{Chair of Econophysics and Complex Systems, \'Ecole polytechnique, 91128 Palaiseau Cedex, France}
\affiliation{LadHyX UMR CNRS 7646, \'Ecole polytechnique, 91128 Palaiseau Cedex, France}
\affiliation{Capital Fund Management, 23 Rue de l’Universit\'e, 75007 Paris, France}

\author{Damien Challet}
\email{damien.challet@centralesupelec.fr}
\affiliation{Laboratoire de Math\'ematiques et Informatique pour la Complexit\'e et les Syst\`emes, CentraleSupélec, Universit\'e Paris-Saclay, 91192 Gif-sur-Yvette Cedex, France}

\date{\today}

\begin{abstract}
We propose a minimal model of the secured interbank network able to shed light on recent money markets puzzles. We find that excess liquidity emerges due to the interactions between the reserves and liquidity ratio constraints; the appearance of evergreen repurchase agreements and collateral re-use emerges as a simple answer to banks' counterparty risk and liquidity ratio regulation. In line with prevailing theories, re-use increases with collateral scarcity. In our agent-based model, banks create money endogenously to meet the funding requests of economic agents. The latter generate payment shocks to the banking system by reallocating their deposits. Banks absorbs these shocks thanks to repurchase agreements, while respecting reserves, liquidity, and leverage constraints.  The resulting network is denser and more robust to stress scenarios than an unsecured one; in addition, the stable bank trading relationships network exhibits a core-periphery structure. Finally, we show how this model can be used as a tool for stress testing and monetary policy design.
\end{abstract}

\maketitle


\section{Introduction}

\subsection{Motivation}

Money markets are the place where banks conduct their refinancing operations. They serve as the engine of the money creation process which provides liquidity to the financial system, thus contributing to its stability. Following the surge in counterparty risk during the 2008's Great Financial Crisis (GFC), money markets in western countries have undergone significant transformations. In October 2008, the European Central Bank (ECB) introduced the so-called full allotment procedure, which allows banks to request unlimited central bank funding. Concurrently, the implementation of the Basel regulation regarding the Liquidity Coverage Ratio (LCR) aims to enhance the short-term resilience of banks to a liquidity crisis. It requires banks to maintain an adequate level of high-quality liquid assets to fulfill their liquidity needs under a stress scenario. These measures have contributed to the emergence of excess reserves in the financial system \citet{Renne-2012,PiquardSalakhova-2019,LucaBaldoEtAl-2017}. Additionally, the refinancing of the banking system has shifted towards collateralized lending (using repurchase agreements, or \textit{repos} further defined below), and the practice of collateral re-use has become increasingly prevalent \citep{KellerEtAl-2014,EuropeanSystemicRiskBoard.-2017,CheungEtAl-2014,FuhrerEtAl-2016,SCAGGS-2018,Accornero-2020}. The network structure of money markets, where transactions among banks are identified as links between nodes, has evolved in consequence. While the unsecured market experiences very low density \citep{BechMonnet-2016,BlasquesEtAl-2018,Vari-2020,BossEtAl-2004}, we observe that repo markets demonstrate higher density due to longer transaction maturities.

Several authors have proposed explanations for recent changes within money markets \citep{Renne-2012, PiquardSalakhova-2019,LucaBaldoEtAl-2017,FilippoEtAl-2018, Allen-2016,JankEtAl-2021, Vari-2020, DubecqEtAl-2016}. However, these approaches generally refer to complex mechanisms that are difficult to quantify, for example, the high opportunity cost to not hold a non-risky coupon \citep{PiquardSalakhova-2019}, market fragmentation \citep{Vari-2020} or collateral scarcity \citep{JankEtAl-2021}. Moreover, the literature on agent-based models (ABM) has focused so far on the absorption of payment shocks by the banking system through unsecured transactions under reserve constraints \citep{Poole-1968,BechMonnet-2016,BlasquesEtAl-2018,Lux-2015,LiuEtAl-2020}. In fact, banks endogenously produce money through lending \citep{JakabKumhof-2015,JakabKumhof-2018} and the use of secured transactions raises non-trivial stability questions.

Here we consider money creation and payment shocks within collateralized markets, subject to reserve, LCR, and leverage constraints. Our model shows that excess liquidity and re-use can be explained by regulatory constraints and repo contracts specificities. Our ABM also generates a trading network with high density, stable bilateral trading relationships, asymmetric in-- and out-- degree distributions as well as a core-periphery structure. Finally, this model is a useful tool for simulating the systemic effects on financial stability of crisis scenarios or regulatory changes.

After describing recently established stylized facts in money markets, we review the existing literature on interbank network modeling. Then, section~\ref{Agent based model description} introduces our ABM. Finally, section~\ref{Results} presents the dynamical behavior of the model, the effects of parameter changes, and stress scenarios.

\subsection{Money markets stylized facts} \label{Money markets stylized facts}

This section presents the stylized facts observed in money markets, which our agent-based model successfully reproduces. In addition to reviewing the empirical literature, we offer our own explanations for some of these phenomena.

\subsubsection{Excess liquidity and declining unsecured interbank markets}

Following the GFC, the ECB implemented a so-called full allotment procedure, which accommodates any liquidity demand from banks in unlimited amounts, as documented by \citet{Renne-2017}. This central bank liquidity is provided in exchange for low-quality collateral, which is abundant in banks' balance sheets. Subsequently, the volumes of the overnight unsecured interbank market have decreased significantly. Indeed, the volumes in interbank secured markets drop when excess reserves (i.e. surplus from bank reserves requirements) increase \citep{PiquardSalakhova-2019}. As explained by a recent ECB survey \citep{LucaBaldoEtAl-2017}, the increase in excess liquidity between $2012$ and $2018$ was mainly driven by (i) the greater demand of banks for central bank liquidity, (ii) the full allotment procedure, and (iii) the offer of longer-term refinancing operations. Since $2015$, another ingredient has led to a new increase in excess liquidity: The ECB has injected central bank liquidity into the banking system through its asset purchase program (APP). This time, most banks cited increasing client inflows as the main reason for their excess liquidity \citep{LucaBaldoEtAl-2017}. The resulting decline in unsecured lending was reinforced by the introduction of the LCR in January $2018$, which hampered the redistribution of liquidity (\citet{LucaBaldoEtAl-2017} and section~\ref{Creation of evergreen repos to answer to the LCR regulation}). 

Using individual bank balance sheets, one can show that the combination of the APP and LCR constraint leads to excess liquidity in the financial system. For a bank $i$ at time $t$, let us denote, respectively, by $C_{i}(t)$ and $D_{i}(t)$, the cash owned by a bank and the deposits it received. We also denote by $S^u_{i}(t)$ its amount of securities usable as collateral and by $S^c_{i}(t)$ the collateral it received as the lender of cash. The LCR is the ratio of unencumbered assets to net cash outflows over the next 30 days. These outflows are defined as a regulatory prescribed haircut of a bank's liabilities. Formally, within the simplified bank balance sheet we defined, the LCR is expressed as
\begin{equation} \label{eq:LCR}
    LCR_{i}(t) = \frac{C_{i}(t)+ S^u_{i}(t) + S^c_{i}(t)}{\beta D_{i}(t)} \geq 100\%,
\end{equation}
where $\beta$ is the regulatory outflow rate for deposits. The excess liquidity~$E(t)$ in the banking system at time~$t$ is defined as the sum of the cash in excess of the minimum reserves of the individual banks:
\begin{equation} \label{eq:def_excess_liquidity}
    E(t) := \sum_{i=1}^N C_{i}(t) - \alpha D_{i}(t),
\end{equation}
where $\alpha$ is the share of minimum reserves required by the regulation. If we replace $C_{i}(t)$ by its expression in equation~\ref{eq:LCR} and assume all banks cover the same outflow rate $\beta$, equation~\ref{eq:def_excess_liquidity} can be written as:
\begin{equation} \label{eq:excess_liquidity}
    E(t) \geq (\beta - \alpha) D(t) - S(t),
\end{equation}
where $S(t)$ and $D(t)$ are respectively the total amount of collateral and deposits in the banking system. Eq.~\eqref{eq:excess_liquidity} shows that the larger the gap $\beta- \alpha$ between the regulatory outflow rate and the required minimum reserve, the higher the excess liquidity. In addition, a decrease in the amount of collateral available generates additional excess liquidity. However, this reasoning does not hold in the presence of interactions between banks. Our ABM actually shows that the asymmetric response of banks to payment shocks also generates excess liquidity even when there is no collateral scarcity (see section~\ref{Banks' behavioral rules}).

\subsubsection{Evergreen repos to answer LCR regulation} \label{Creation of evergreen repos to answer to the LCR regulation}

The GFC highlighted the existence of counterparty risk among banks. This led to a transition from unsecured to secured lending \citep{FilippoEtAl-2018}. Within these markets, collateralized borrowings are performed thanks to repos, i.e., financial contracts exchanging collateral against cash for a given time period. In fact, the substitution effect towards secured markets was reinforced by the introduction of the LCR because of the ability of such contracts to circumvent this constraint. A repo contract continuously renewed by mutual agreement is called an evergreen repo. We show in the following that an evergreen repo with a one-month notice period has no effect on the LCR of the two involved parties. Empirical evidences show that the introduction of the LCR regulation coincides with an increase in the volumes of traded evergreen repos with a notice period of more than a month \citep{Allen-2016,LeCozEtAl-2024b}.  In particular, \citet{LeCozEtAl-2024b} observed that the volume of evergreen repos traded among the $50$ largest banks in the eurozone increased from negligible amounts in $2017$ to ten billions per day in $2019$.

The use of evergreen repos to circumvent LCR regulation can be explained thanks to the simplified balance sheet of a bank. In the context of a repo agreement, the borrower of cash $i$ remains the owner of the collateral he provided for the transaction. This collateral remains on its balance sheet as \textit{encumbered} securities denoted by $S^e_{i}(t)$. A new repo of notional $\Delta R$ leads to the increase of encumbered collateral to $S^e_{i}(t) + \Delta R$. This collateral cannot be used in any other transaction, thus it is excluded from the numerator of the LCR. In contrast, the lender $j$ of cash records this collateral as received collateral $S^c_{i}(t)$, which compensates for his loss of LCR due to its cash reduction. After the transaction, at $t+\Delta t$, the new LCR of the borrower $i$ and the lender $j$ remain constant:
\begin{align} \label{LCR conservation}
    LCR_{i}(t+\Delta t) & =\frac{(C_{i}(t)+\Delta R)+ (S^u_{i}(t)-\Delta R) + S^c_{i}(t)}{\beta D_{i}(t)} \nonumber\\
    &=  LCR_{i}(t), \\
    LCR_{j}(t + \Delta t) & = \frac{(C_{j}(t)-\Delta R)+S^u_{j}(t) + (S^c_{j}(t)+\Delta R)}{\beta D_{j}(t)} \nonumber \\
    &=  LCR_{j}(t).
\end{align}
LCR's denominator is the total outflow, generated during a one-month stress test, by a bank's liabilities. In equation~$\ref{LCR conservation}$, one could be surprised not to see any outflow from the repo recorded as a liability for the cash borrower. This is possible only if we consider a repo with maturity (or notice period) greater than one month. In addition, to ensure the LCR conservation equation~$\ref{LCR conservation}$ for all time $t$, it is necessary to introduce \textit{evergreen} contracts valid at all times. On the opposite, entering in a unsecured interbank loan $\Delta U$ would negatively impact the LCR of the lender and positively impact the LCR of the borrower:
\begin{align} \label{LCR non-conservation}
    LCR_{i,t+\Delta t} & =\frac{C_{i}(t)+\Delta U+ S^u_{i}(t) + S^c_{i}(t)}{\beta D_{i}(t)} \geq  LCR_{i}(t), \\
    LCR_{j,t + \Delta t} & = \frac{C_{j,t}-\Delta U+S^u_{j,t} + S^c_{j,t}}{\beta  D_{j,t}} \leq  LCR_{j,t}.
\end{align}

The substitution effect between the unsecured and secured markets is also influenced by the APP, which (i) increases the spread between the secured and unsecured rates due to the lower availability of the collateral, and (ii) decreases the volumes on both the secured and unsecured markets as a consequence \citep{PiquardSalakhova-2019}. According to these authors, the decline in volumes stems from collateral scarcity in the secured segment, while the unsecured segment appears too costly due to a higher spread relative to the secured one. More intuitively, as demonstrated in our model, the APP increases excess liquidity, thereby reducing the need for interbank funding.

\subsubsection{Collateral re-use and bond scarcity}
The one-month notice period of evergreen repo forbids the immediate unwinding of existing positions when a lender of cash experiences a liquidity need. Thus, these markets offer the possibility to \textit{re-use} collateral: the lender of cash $j$ is allowed to re-use the collateral $S^c_{j,t}$ he received during a reverse repo in order to borrow cash within another repo transaction. Various definitions have been used to define the re-use rate of collateral within money markets \citep{Accornero-2020}. Here we choose the following definition:
\begin{align}
    \text{re-use}(t) = \frac{\sum_{i=1}^NS_i^r(t)}{\sum_{i=1}^NS^c_i(t)},
\end{align}
where $S_i^r(t)$ is the collateral re-used by bank~$i$ in the context of a repo. Various levels of collateral re-use, ranging from $0.1$ to $3$ have been measured across time and regions: notably a re-use rate around $1$ was observed in European money markets \citep{KellerEtAl-2014, EuropeanSystemicRiskBoard.-2017, LeCozEtAl-2024b}, $0.6$ in Australia \cite{CheungEtAl-2014}, $0.1$ in Switzerland \citep{FuhrerEtAl-2016}, and $3$ in the US \citep{SCAGGS-2018}. The high re-use rate observed on money markets is not a threat to the initial objective of the LCR regulation. Indeed, the same collateral can only appear once at the numerator of the LCR of a given bank. The other appearances of this collateral are identified as encumbered securities, which are excluded from the LCR.

Re-use increases in response to the scarcity induced by the APP \citep{JankEtAl-2021}. Moreover, re-use contributes to the buildup of leverage \citep{Theinternationalcapitalmarketassociation-2015, BrummEtAl-2018, HorenKotidis-2018} by inflating balance sheet sizes. Using an infinite-horizon asset-pricing model with heterogeneous agents, \citet{BrummEtAl-2018} considers that this increased leverage then significantly increases volatility in financial markets, ultimately reducing welfare.  

\subsubsection{The interbank network topology} \label{The interbank network topology}

\paragraph{Sparse core periphery structure?} We define a link in the interbank network as the existence, over a given aggregation period (typically ranging from one day to one year), of at least one repo exposure between two banks. Historically, interbank market networks have been characterized by a low density and a core-periphery structure \citep{BechMonnet-2016,BlasquesEtAl-2018,Vari-2020, BossEtAl-2004}. In this network configuration a central 'core' of highly interconnected nodes is surrounded by a 'periphery' of less connected nodes that primarily connect to the core rather than to each other. The switch of these markets towards secured transactions led to an increased network density \citep{LeCozEtAl-2024b}. These authors show a network density ranging from $10\%$ to $20\%$ depending on link definition conventions. We assume that this higher density is due to the longer transaction maturity. The limited number of banks in our sample (only the 50 largest ones out of the 3000 financial institutions in the eurozone) prevented us from observing a core-periphery structure of secured markets.

\paragraph{Stable bilateral relationships}
The existence of stable interbank relationship lending has been documented, among others, by \citet{Furfine-1999,AfonsoEtAl-2013,BlasquesEtAl-2018, LeCozEtAl-2024b}. In the case of secured markets \citet{LeCozEtAl-2024b} measured the share of stable links from one period to another, namely the Jaccard network similarity index \citep{VermaAggarwal-2020}, ranging from $80$ to $100\%$ depending on the aggregation period defining links.

\paragraph{Asymmetric in and out degrees?}
Several authors reported an asymmetry between in and out degree within unsecured interbank lending networks \citep{CraigvonPeter-2014,AnandEtAl-2015,Lux-2015}. Notably, \citet{CraigvonPeter-2014,AnandEtAl-2015} observe that banks in Germany have typically (according to the median) fewer lenders than borrowers. \citet{LeCozEtAl-2024b} observe a more symmetrical pattern in the case of the repo exposures among the 50 largest banks of the eurozone.

\FloatBarrier
\subsection{Money markets modeling in the literature}

Several approaches to the modeling of interbank unsecured markets have been proposed. The influential article of \citet{Poole-1968} introduces a model in which the interbank lending network absorbs randomly generated payment shocks, under reserves requirements’ constraints. This seminal work has been followed by numerous proposals of network modeling of the interbank market. Recently, \citet{HeiderEtAl-2015} included counterparty risk in the lending network and generated endogenous liquidity hoarding. \citet{BechMonnet-2016} considered a search-based model that can reproduce the decrease in trading volumes due to a surge in excess reserves, without identifying the initial cause of deposits surpluses. This was later identified by \citet{Vari-2020} as the eurozone interbank market fragmentation: banks, depending on their country of location, have different probabilities of default. This fragmentation disrupts the transmission of monetary policy, generating endogenously excess liquidity. \citet{Vari-2020} distinguishes two groups: core banks (in Germany and the Netherlands) do not use central bank funding but hold excess reserves; peripheral banks (e.g., in Spain and Italy) borrow massively from the central bank to fulfill their needs. However, the funding obtained ends up within the core banks due to payment imbalances. We reproduce the same behavior: in our ABM, the flow of payment shocks moves deposits from peripheral to core banks, generating liquidity needs in the first ones, and excess liquidity in the others (see section~\ref{Banks' behavioral rules}). More recently, the decline of unsecured markets led several authors \citep{PiquardSalakhova-2019,DeFioreEtAl-2021} to build equilibrium models explaining the substitution effects between secured and unsecured interbank markets. However, all these modeling proposals can only describe the equilibrium state of the interbank market.

Other authors proposed dynamic models of unsecured interbank markets \citep{AfonsoLagos-2015, BlasquesEtAl-2018, LiuEtAl-2020, Lux-2015, Halaj-2018}. In particular, \citet{BlasquesEtAl-2018} assume profit maximization and risk monitoring cost to generate a sparse core-periphery structure and stable bilateral trading relationships. \citet{Lux-2015} obtain the same result using a reinforcement-learning scheme. \citet{LiuEtAl-2020} proposed an ABM of the interbank network that leads to the endogenous formation of a financial network using only data from individual banks.

Within ABM models for Macroeconomics, several frameworks include multiple bank agents from which firms can borrow, although they do not allow interactions among banks \citep{DawidEtAl-2012, DawidEtAl-2018, DawidGemkow-2014, DawidEtAl-2016, CincottiEtAl-2012, DosiEtAl-2010, DosiEtAl-2013, DosiEtAl-2015, DosiEtAl-2017, RiccettiEtAl-2013,BargigliEtAl-2014,RiccettiEtAl-2015,RiccettiEtAl-2018,BargigliEtAl-2020,RiccettiEtAl-2022}. The modeling of a static interbank lending market is found in some macroeconomic models \citep{SchasfoortEtAl-2017, Reissl-2018, Reale-2019, GurgoneEtAl-2018}.

Overall, these approaches focus on unsecured markets, which have been largely replaced by secured markets. Moreover, these models assume the absence of endogenous money creation while this process induces non-centered shocks requiring a specific modeling \citep{JakabKumhof-2015, JakabKumhof-2018}. Finally, these frameworks only account for reserves constraints while the introduction of the LCR significantly modified money markets (see section~\ref{Money markets stylized facts}). 

\section{A minimal agent-based model} \label{Agent based model description}
We consider a money market formed by $N$ bank agents, a representative economic agent, and a central bank. Banks can create money by lending to the economic agent. The latter then reallocates his deposits among bank agents, thus generating payment shocks. These shocks are absorbed by the banking system thanks to central bank funding and repos. We assume the existence of a single type of fungible security usable as collateral in the repo market, typically a government bond. Banks are also assumed to access unsecured funding from the central bank. While such funding is typically collateralized, the eligible collateral is of lower quality than that used in interbank markets. We model the limiting case where this low-quality collateral is so abundant that all banks effectively hold it in unlimited amounts—rendering funding effectively unsecured. To our knowledge, no scarcity of such collateral has been observed. Modeling two collateral tiers within an agent-based framework is left for future work. Finally, bank agents must respect at all times their reserves, LCR, and leverage regulatory requirements. None of the fixed income instruments in the system offers any coupon.

\subsection{Balance sheet items} \label{Balance sheet items}
Each bank $i$ is characterized by the following accounting items, expressed in monetary units, at each time step $t$ (in units of day).
\begin{itemize}[leftmargin=0.4cm]
    \item Assets:
    \begin{itemize}[leftmargin=+0.1cm]
        \item Cash: either deposits at the central bank or reserves, denoted by $C_{i}(t)$;
        \item Securities usable as collateral, denoted by $S_i^u(t)$;
        \item Securities encumbered in the context of a repo, denoted by $S_i^e(t)$;
        \item Loans to the economic agent denoted by $L_{i}(t)$;
        \item Reverse repos granted to other banks, $$R^r_{i}(t) =\sum_{j\neq i} r^r_{i,j}(t),$$ where $r_{j,i}(t)$ denotes the sum of the open repo exposures at time $t$ that were received by the bank $i$ from the bank $j$.
    \end{itemize}
    \item Liabilities:
    \begin{itemize}[leftmargin=+0.1cm]
        \item Own funds or equity, $O_{i}(t)$;
        \item Deposits, $D_{i}(t)$;
        \item Repo exposures received from other banks, $$R_{i}(t) =\sum_{j\neq i} r_{i,j}(t);$$
        \item Central bank funding, denoted by $M_{i}(t)$;
    \end{itemize}
    \item Off-balance sheet:
    \begin{itemize}[leftmargin=+0.1cm]
        \item Collateral received in the context of a reverse repo, denoted by $S_i^c(t)$;
        \item Collateral re-used in the context of a repo, denoted by $S_i^r(t)$.
    \end{itemize}
\end{itemize}

\subsection{Financial contracts}
The financial contracts in the model can have either an infinite maturity or no maturity. Repos are evergreen (i.e. have unlimited maturity) with a one-month notice period for cancellation. Therefore, banks must create new repos to remediate immediate liquidity needs as the unwinding of existing reverse repos would provide liquidity too late. When a bank is in excess of cash, it would also have to wait $30$ days to unwind its existing repo, while it could immediately earn the repo rate when entering a new reverse repo. Loans, central bank funding, and securities have unlimited maturity. Deposits and cash have no maturity.

As mentioned above, we assume that none of these financial instruments offers any coupon. Indeed, simulating yields dynamics is not necessary to reproduce excess liquidity, repo re-use, and network topology stylized facts. However, the behavioral rules attributed to banks in our model are meaningful only if a relative ordering of the yields across financial contracts is also assumed. These rules are then justified by the principle that banks prefer holding the instrument delivering the highest coupon. This requires defining the relative static yields of each financial contract. Hence, in our model, securities used as collateral deliver a higher interest rate than the discount facility rate remunerating banks' cash balances. This assumption is consistent with empirical observations. For example, in the eurozone, $10$-year German government bonds have almost systematically delivered higher coupons than the ECB discount facility rate. In addition, the rate of the central bank funding is higher than the repo rate, therefore banks have an incentive not to borrow from the central bank. The repo market rate is higher than the discount rate, so banks accept entering into reverse repo when they are in excess of cash. Finally, we assume that the loan rate to the real economy is the highest rate available to a bank agent.

\subsection{Regulatory constraints}
Banks are subject to three regulatory obligations.
\begin{enumerate}
    \item The minimum reserves constraint:  banks must keep a share of the deposits they receive in the form of central bank reserves, i.e.,
    \begin{align}
        C_{i}(t) \geq \alpha D_{i}(t).
    \end{align}

    \item The LCR constraint, requiring banks to maintain the ratio of their unencumbered assets to cash outflows over the next 30 days above one; within our model's balance sheet for banks, the LCR constraint amounts to
    \begin{align}
    \label{eq:LCR_constraint}
        C_{i}(t)+ S^u_{i}(t) + S^c_{i}(t) \geq \beta D_{i}(t),
    \end{align}
    assuming a regulatory net deposit outflow $\beta$ and that the securities received in the context of a repo~$S^c_{i}(t)$ will remain unencumbered during a one-month stress test. In the following sections, we will refer to the effective $\beta_i(t)$ defined by
    \begin{align}
        \beta_i(t) = \frac{C_{i}(t)+ S^u_{i}(t) + S^c_{i}(t)}{D_{i}(t)},
    \end{align}
    as the liquidity ratio or the LCR ratio of the bank $i$. This means that the bank can face an outflow rate $\beta_i(t)$ of its deposits -- which must be higher than the regulatory $\beta$.

    \item The leverage ratio (or solvency ratio) constraint, requiring banks to keep their own funds above a certain share of their total assets:
    \begin{align}
        O_{i}(t) \geq \gamma\left(C_{i}(t)+S^u_{i}(t) +S^e_{i}(t) +L_{i}(t) + R^r_{i}(t)\right).
    \end{align}
\end{enumerate}
It is worth mentioning that the leverage ratio plays the same role as the solvency ratio, as it requires banks to maintain a minimum level of own funds. The solvency ratio is more complex to account for as it involves risk measurement. It is also less binding than leverage constraints for low risk activities \citep{BourahlaEtAl-2018}. Thus, we choose to ignore solvency ratio constraints in our model.

\subsection{Initialization or money creation}
All financial instruments in the model are created endogenously. Each bank $i$ can create an amount $\Delta X_{i}(t)$ of new money at step $t$ by lending cash to the representative economic agent. The latter must then store the same amount in the form of a deposit at the bank $i$. To ensure that the money creation process is compatible with the three regulatory constraints, the value of newly created securities and own funds must be proportional to that of new loans. Securities are typically government bonds issued by the representative economic agent and bought by the banking system. As the government also stores the borrowed cash in the form of a deposit to the banking system, this mechanism increases the usable deposits and securities in the banks' balance sheets. In addition, own funds are issued by banks and bought by the economic agent using some of the cash borrowed from banks.

In summary, the creation of $\Delta X_{i}(t)$ monetary units by the bank $i$ at step $t$ involves three steps: (i) lending, (ii) issuance of government bonds, and (iii) capital increase of bank $i$ by issuing new shares. The combined effect of this three actions results in the increase of each of the balance sheet item $A$ by its corresponding variation  $\Delta A$:
\begin{equation}
    \begin{cases}
    \label{eq:money_creation_shock}
        \Delta D_{i}(t) := (1-\gamma_{\text{new}})\Delta X_{i}(t) \\
        \Delta L_{i}(t) := \Delta X_{i}(t) - \beta_{\text{new}}(1-\gamma_{\text{new}})\Delta X_{i}(t) \\
        \Delta S^{u}_{i}(t) := \beta_{\text{new}}(1-\gamma_{\text{new}})\Delta X_{i}(t) \\
        \Delta O_{i}(t) := \gamma_{\text{new}} \Delta X_{i}(t),
    \end{cases}
\end{equation}
where $\gamma_{\text{new}} \in [0,1]$ and $\beta_{\text{new}} \in [0,1]$ are the parameters governing respectively the issuance of shares and securities. In practice, unless otherwise specified, we assume $\beta_{\text{new}} = \beta$, such that enough collateral is created to meet regulatory obligations. We study a case where $\beta_{\text{new}} \neq \beta$ in the section~\ref{Stress testing} when we simulate the effect of an APP. In addition, we consistently assume $\gamma_{\text{new}} > \gamma$, which means that banks can issue sufficient own funds to meet their leverage constraints. If this assumption were relaxed, banks would quickly become unable to generate new loans. The other accounting items are generated either by (i) repo transactions (encumbered securities, collateral received, and collateral re-used) or (ii) central bank funding (main refinancing operations and cash).

\subsection{Money creation shocks} \label{Money creation shocks}
We simulate money creation thanks to a multiplicative random growth process in which shocks fluctuate around an average rate $g$ of new money. Let $(Z_i(t))$ be log-normal random variables of volatility $v$ independent across banks $i$ and steps $t$. The amount of created money $\Delta X_i(t) = X_i(t+1)-X_i(t)$ is given by
\begin{align}
\label{eq:money_creation}
    &\Delta X_i(t) = g Z_i(t) X_i(t), \nonumber \\
    &X_i(0) = x_0 Z_i(0),
\end{align}
where $g$ is the growth rate of money.

Neither the process $X_i(t)$ nor its normalized version $\frac{X_i(t)}{\sum_{i=0}^N X_i(t)}$ converge towards a stationary distribution \citep{MarsiliEtAl-1998,Gabaix-1999,Mitzenmacher-2004,BouchaudMezard-2000}.  However, we report in appendix~\ref{Random growth model} that the normalized size of banks $\frac{X_i}{\sum_i X_i}$ behaves as a non-stationary log-normal distribution that evolves very slowly compared to the typical time scale of the model. Notably, the tail of this log-normal distribution remains stable within a given range, for a sufficiently long time (around $5000$ steps) for the network to reach a state close to stationarity (see section~\ref{Typical run}). It is not feasible to design a random growth model that generates a stationary limit using the approaches proposed by \citet{MarsiliEtAl-1998,Gabaix-1999, BouchaudMezard-2000}. Indeed, these models either require defining a negative drift \citep{MarsiliEtAl-1998,Gabaix-1999} or facilitating cash exchanges between banks \citep{BouchaudMezard-2000} (see appendix~\ref{Random growth model}).

In the empirical literature, there is no consensus regarding the size distribution of banks. Most authors \citep{Lux-2015,JanickiPrescott-2006,CerquetiEtAl-2022} suggest that this distribution follows a power law with a tail exponent between $1$ (Zipf's law) and $3$ across time and regions. However, \citet{GoddardEtAl-2014} argue that bank sizes are better described by a truncated log-normal distribution. Differentiating between a power law and a log-normal distribution is challenging with small sample sizes. In the context of banks, there are only a few thousand financial institutions within a given monetary zone, which limits the ability to accurately assess their size distribution.

In our model, as long as bank sizes are sufficiently heterogeneous, we observe that the specific distribution of bank sizes (log-normal or power law) does not influence the stylized facts previously mentioned. Hence, in order to reach faster stationarity, we conduct our parameter space (see section~\ref{Parameter space}) and stress tests analyses (see section~\ref{Stress testing}) by initializing money creation $X_i(0)$ as a power law of tail exponent $\nu$. In such a case, the volatility $v$ of the random growth is set to zero to maintain the initial size distribution of banks over time.

\subsection{Payment shocks}
Once money is created, economic agents transact goods. Each transaction results in an increase in the deposits in the bank of the seller and a decrease in the deposits in the bank of the buyer. The total amount of deposits in the banking system remains constant during these transactions. Similarly to the approach of \citet{Lux-2015}, we simulate payments thanks to normally distributed shocks defined to ensure that (i) the total sum of deposits is conserved and (ii) there is mean reversion toward the amount of deposits created by the bank. Formally, the deposits variation caused by payment shocks at step $t$ for the bank $i$ is defined by
\begin{align}
    \Delta' D_{i}(t):=\sigma & \Bigg[\Bar{D}_{i}(t)-D_{i}(t)+\epsilon_{i}(t) D_{i}(t) \nonumber \\
   & - \bigg(\frac{1}{N} \sum_{j=1}^N \Bar{D}_{j}(t)-D_{j}(t)+\epsilon_{j}(t) D_{j}(t)\bigg) \Bigg],
\end{align}
where $(\epsilon_{i}(t))$ are normalized centered and independent Gaussian shocks and $\Bar{D}_{i}(t) = (1-\gamma_{\text{new}}) X_{i}(t)$ is the target of the mean reversion, updated according to the money creation process $X_{i}(t)$.

For large values of $\sigma$, it is possible that the deposit shock $\Delta'D_i(t)$ increases in absolute value compared to current bank deposits $i$. To ensure that deposits after the shock (that is, $D_i(t) + \Delta'D_i(t)$) are positive, we choose $\sigma \leq 10\%$. This means that a shock must exceed $10\times\sigma$ to generate negative deposits. Although such events are very rare, we apply a floor to banks' deposits, preventing them from going below zero.

Stock flow consistency imposes to increase the cash balance of bank $i$ at each time step $t$ by the same amount, i.e.,
\begin{align}
    \Delta' C_{i}(t):=\Delta' D_{i}(t).
\end{align}

\subsection{Banks' behavioral rules} \label{Banks' behavioral rules}
Money creation and payment shocks modify the balance sheet of banks and can lead to a breach of their regulatory constraints. If that is the case, central bank funding and repo markets are used by bank agents to meet these obligations. To enhance readability in this section, we assume that all the inequalities characterizing regulatory constraints for bank agent~$i$ are equalities before the money creation and payment shocks, i.e.
\begin{equation}
    \begin{cases}
        C_{i}(t) = \alpha D_{i}(t), \\
        C_{i}(t) + S^u_{i}(t) + S^c_{i}(t) = \beta D_{i}(t), \\ 
        O_{i}(t) = \gamma^* \left(C_{i}(t)+S^u_{i}(t) +S^e_{i}(t) +L_{i}(t) + R^r_{i}(t)\right).
    \end{cases}
\end{equation}
In fact, the model structurally generates excess liquidity (i.e. $C(t) \geq \alpha D(t)$) and excess LCR (i.e. $C_{t} + S^u(t) + S^c(t) \geq \beta D(t)$), because of the asymmetric responses of banks to payment shocks, as further described below. At the beginning of step $t+1$, the bank $i$ receives a money creation shock and a payment shock. To meet its three regulatory constraints, the bank will act as follows.
\begin{enumerate}
    \item \textit{LCR management}. Secured lending keeps the LCR level unchanged (see section~\ref{Creation of evergreen repos to answer to the LCR regulation}). Hence, in the absence of an unsecured market, banks optimize their LCR levels through central bank funding. We denote $\Delta M_{i}(t)$ the amount of central bank funding that the bank $i$ will request or end from to maintain its LCR at the level $\beta$. Bank $i$ must minimize their central bank funding $ M_{i}(t)$ such that: 
    \begin{align}
        \Delta M_{i}(t) \geq \beta \left(\Delta D_{i}(t) + \Delta' D_{i}(t)\right) - \Delta S^u_{i}(t) - \Delta' D_{i}(t).
    \end{align}
    This equation is derived from all the terms  in equation~\eqref{eq:LCR_constraint} that change following a money creation and payment shocks, while replacing $\Delta' C_i(t)$ by $\Delta' D_i(t) + \Delta M_i(t)$.  We assume that the share of securities created during the money creation process is equal to the regulatory LCR level $ \beta_{\text{new}} = \beta$, thus, according to Equation~\eqref{eq:money_creation_shock}, $\Delta S^u_i(t)=\beta \Delta D$. Hence, a negative payment shock will lead the bank to request $-\big(1-\beta\big)\Delta' D_{i}(t)$ of central funding. In case of a positive shock, the only strategy for bank~$i$ to reduce its LCR is to reimburse its existing central bank funding $M_i(t)$. If the existing central bank funding is lower than $-\big(1-\beta\big)\Delta' D_{i}(t)$, the bank will remain is excess of LCR. Overall, the optimal funding is given by
    \begin{align}
        \Delta M_{i}(t) =  \max{\left\{-\big(1-\beta\big)\Delta' D_{i}(t),-M_{i}(t)\right\}}.
    \end{align}
    According to this behavioral rule, the net sum of central bank funding across all banks is positive, which introduces excess liquidity in the system.
    
    \item \textit{Reserve management}. Banks use repos to optimize their central bank reserves. The one-month notice period of these contracts requires banks to open new long or short positions to manage their short-term liquidity. Banks will close some of their existing repos only to meet their leverage ratio obligations (see next paragraph). We denote $\Delta R_{i}(t)$ the amount of repo requested by the bank $i$ (if $\Delta R_{i}(t)>0$) or of reverse repo that the bank $i$ is willing to accept (if $\Delta R_{i}(t)<0$) in order to maintain its LCR at a target level $\beta$. Bank $i$ must minimize its repo exposure $R_{i}(t)$ to meet its reserves constraints at step~$t+1$:
    \begin{align}
        C_{i}(t) + \Delta' D_{i}(t) + \Delta M_{i}(t) +\Delta R_{i}(t) \nonumber \\
        \geq \alpha \big(D_{i}(t) + \Delta D_{i}(t) + \Delta' D_{i}(t) \big).
    \end{align}
   If $\beta = \beta_{\text{new}}$ and the bank was not in excess of reserve before the shocks, we have
    \begin{align}
        \Delta R_{i}(t) = -\Delta M_{i}(t)  - (1-\alpha) \Delta' D_{i}(t) \nonumber \\ 
        + \alpha \Delta D_{i}(t).
    \end{align}

   If we also assume an absence of excess LCR, i.e. $\Delta M_{i}(t) =  -\big(1-\beta\big)\Delta' D_{i}(t)$, the previous equation becomes
   \begin{align}
        \Delta R_{i}(t) = -\left\{(1-\alpha) - (1-\beta) \right\}\Delta' D_{i}(t) \nonumber \\
        + \alpha \Delta D_{i}(t).
    \end{align}
    Most banking regulations typically set $\alpha < \beta$, so the difference $(1-\alpha) - (1-\beta)$ is positive. If we neglect money creation shocks (i.e. $\Delta D_{i}(t) \ll \Delta' D_{i}(t)$), it is clear that receiving a negative payment shock implies requesting repos. In contrast, a positive shock leads the bank to be willing to enter into a reverse repo. Nevertheless, it is possible that this bank does not hold sufficient collateral to enter into a repo, in this case, it will request additional central bank funding.

    \item \textit{Leverage management}. The management of reserves through the opening of repos and reverse repos inflates banks' balance sheets \citep{Theinternationalcapitalmarketassociation-2015, BrummEtAl-2018, HorenKotidis-2018}. If the current leverage ratio of a bank becomes lower than its targeted level $\gamma^*$, it will start ending its existing repos after each positive shock. Contrary to the LCR and reserves constraints, banks do not have immediate solutions available to reduce the size of their balance sheet. Hence they choose a target leverage ratio greater than the regulatory requirement, $\gamma^*>\gamma$. As a consequence, banks start closing their existing repos before risking a breach of their minimum leverage ratio.
\end{enumerate}
         
\subsection{Sequence of the interactions among agents}
We assume that repos are initiated and ended by the borrowers of cash. Banks are ready to participate in the repo market after the individual management of their LCR. Market clearing is performed as follow:
\begin{enumerate}
    \item All banks having to end existing repos do so one by one in a shuffled order. Bank~$i$ starts by contacting its counterparts with the lowest trust level $\phi_{ij}$ (whose dynamics is described below). If the lender of cash $j$ has not sufficient collateral $S_j^c$ to end its reverse repo, bank $j$ must end some of its own repos. We assume that the lender of cash receives its cash back slightly before providing back the collateral to the cash borrower\footnote{In practice, nothing guarantees the cash to be available before the reimbursement of the collateral, yet, an overnight central bank borrowing would temporarily provide the required liquidity, although it would have no effect at the daily time scale modeled here.}. This mechanism ensures that the lender of cash $j$ owns enough cash to close its existing repos and get back its re-used collateral. This situation can trigger a cascade of collateral call backs (further detailed below).
    \item  Then, banks having to enter into repos do so one by one in another reshuffled order. This time, the bank $i$ starts by contacting their counterparts with the highest trust factor $\phi_{ij}$. These latter accept entering into a reverse repo if they are in excess of cash, i.e. $\Delta R_{j,t}<0$.
\end{enumerate}

Naturally, banks can engage simultaneously into repos and reverse repos. However, this can lead to a collateral loop if a security is loaned to one bank and then re-loaned to the original lender. In such cases, in our model, the sequential call of collateral to unwind existing repos might not converge. Prohibiting all collateral loops would lead to a rapid collapse of money markets because of the high density of the repo network. Therefore, our model permits these loops, even though it means that some simulations may not run to completion. In actual markets, when two counterparties within a collateral loop want to unwind their positions, they compute directly their net exposures. It is possible that none of the two banks still owns the collateral. In this case, the counterparty who is short of collateral would usually borrow the security (using a reverse repo). In our model, as a simplification, only banks experiencing liquidity needs request funding through repos.

\subsection{Trust coefficients}
As proposed by \citet{Lux-2015}, bilateral trading relationships rely on trust coefficients $\phi_{ij} \in [0,1]$ indicating the strength of the ties established by repeated contact. The trust coefficient from the bank $i$ to $j$ is initialized randomly and updated each time $i$ requests a repo from bank $j$. $\phi_{ij}$ increases if $j$ agrees to lend to $i$ and decrease otherwise: $\phi_{ij}(t+1) =$
\begin{align}
    \label{eq:trust}
     \phi_{ij}(t) + \lambda \left(\frac{(\min( \Delta R_{i}(t), -\Delta R_{j,t}))^+}{\Delta R_{i}(t)}-\phi_{ij}(t)\right)
\end{align}
where $\lambda$ is a learning coefficient, governing the time scale at which banks update their trusts and $(x)^+$ is the maximum between $x$ and $0$. In Eq.~\eqref{eq:trust} the term $(\min( \Delta R_{i}(t), -\Delta R_{j,t}))^+$ is the amount of repo that satisfies the needs of the two counterparties. Thus, the trust coefficients converge towards the share of the repo exposure accepted by the bank $j$ at a speed governed by $\lambda$.

\subsection{Synthesis}
We can sort the variables and parameters of the model into four categories:
\begin{enumerate}
    \item Four exogenous variables are set by the economic agent through monetary and payment shocks: the amount of deposits $D$, loans $L$, securities $S^u$, and own funds $O$.
    \item 7 parameters act as the control parameters of the model. They are constant across banks and time. $\beta_{\text{new}}$, the deposit outflow rate equivalent of new securities, tunes the creation of new securities in the system. $\gamma^*$, the target leverage ratio, and $\gamma_{\text{new}}$, the leverage ratio equivalent of new own funds, control the repo re-use rate.  $g$ and $v$ are respectively the mean and volatility of monetary shocks.  If specified, the exponent $\nu$ of the power law distribution of bank sizes governs the heterogeneity among banks. $\sigma$ is the volatility of payment shocks size and $\lambda$ controls the speed at which trust levels are updated.
    \item The regulatory constraints are set by the regulator.
    \item The other variables are endogenously updated.
\end{enumerate}
Table~\ref{tab:variables} provides the list of variables defining and controlling the state of the bank agent $i$.
\begingroup
\begin{table}[h]
    \begin{ruledtabular}
    \begin{tabular}{p{0.1\linewidth} p{0.60\linewidth} p{0.30\linewidth}}
    \multicolumn{1}{c}{\,} & \multicolumn{1}{l}{Definition}  & \multicolumn{1}{l}{Type} \\
    \hline
    $L_i$ &  Loans & Exogenous  \\
    $D_i$ & Deposits & Exogenous  \\
    $\gamma^*$ & Target leverage ratio & Control  \\
    $\beta_{\text{new}}$ & Deposit outflow rate equivalent of new securities & Control  \\
    $\gamma_{\text{new}}$ & Leverage ratio equivalent of new own funds & Control  \\
    $v$ & Volatility of money creation shocks & Control  \\
    $g$ & Mean growth of money creation & Control  \\
    $\sigma$ & Volatility of payment shocks & Control  \\
    $\lambda$ & Learning coefficient to update trust & Control  \\
    $\phi_{ij}$ & Trust to the bank $j$ & Endogenous  \\
    $C_i$ &  Cash account & Endogenous \\
    $S^u_i$ &  Securities usable & Endogenous \\
    $S^e_i$ &   Securities encumbered  & Endogenous\\
    $R^r_i$ &  Reverse repos granted to other banks & Endogenous  \\
    $O_i$ &  Own funds & Endogenous  \\
    $R_i$ & Repo exposures & Endogenous  \\
    $M_i$ & Central bank funding & Endogenous \\
    $S^c_i$ & Collateral received & Endogenous  \\
    $S^r_i$ & Collateral re-used & Endogenous  \\
    $\alpha$ & Regulatory share of minimum reserves & Regulation \\
    $\beta$ & Regulatory LCR outflow of deposits & Regulation  \\
    $\gamma$ & Regulatory leverage ratio & Regulation  \\
    \end{tabular}
    \end{ruledtabular}
    \caption{List of parameters and variables. \label{tab:variables}}
\end{table}
\endgroup
\FloatBarrier

\section{Results} \label{Results}
\subsection{Dynamical behavior} \label{Typical run}
A typical run of the model reproduces most of the money markets' stylized facts in its stationary state. Unless specified differently we fix $g=0.04\%$, and $v=5$. It means banks increase their balance sheet by $10\%$ a year ($1$ year $\approx 250$  steps) on average but some agents can grow on a given day at a $50\%$ annualized rate. We also set $N=300$ to obtain results comparable to those from MMSR database containing $50$ banks \citep{LeCozEtAl-2024b}.  Payment shocks are assumed to be quite volatile ($\sigma=5\%$). All regulatory parameters are set according to their actual value in the euro zone $\alpha=1\%$, $\beta_{\text{new}}=\beta =50\%$ (that is, the average outflow rate for all types of client deposits), $\gamma=3\%$. We choose $\gamma^*=4.5\%$ to ensure that banks satisfy their leverage constraints and $\gamma_{\text{new}}=9\%$ to generate sufficient collateral re-use. We also set the learning coefficient $\lambda$ to $0.5$ and the average initial size of banks $x_0$, to $0.01$ monetary units. Indeed, we assume that one monetary unit in the model corresponds to one billion euros. Thus, we initialize the average capital of all banks to $10$ millions euros, which is close to the minimal own funds required for a banking license ($5$ millions in the eurozone, article 12 of CRD IV).

\begin{figure}
\centering
 \includegraphics[width=\linewidth]{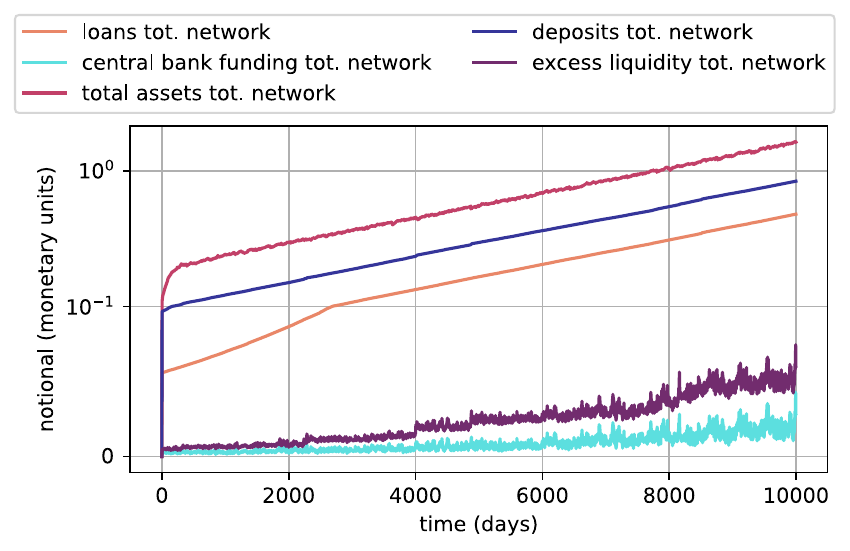}
 \caption{Time evolution of the main macroeconomic aggregates in the simulated banking system with $N=300$ banks, $g=0.04\%$, $v=5$, $\sigma=5\%$, $\beta_{\text{new}}=\beta=50\%$, and $\gamma_{\text{new}}=9\%$.}
\label{fig:typical_run/macro_economic_aggregates}
\end{figure}

Excess liquidity naturally appears as the result of asymmetric responses to payment shocks (see Fig.~\ref{fig:typical_run/macro_economic_aggregates}). The amount of excess liquidity generated by the model is between $5$ and $10\%$ of total assets, in line with the levels observed in the eurozone \citep{HudepohlEtAl-2024}.  This shows that the origin of excess liquidity can be traced back to the interactions between the reserves and LCR constraints, as banks cannot maintain both requirements to their minimum levels and absorb daily payment shocks. Note that the exponential growth of the banking system at a $10\%$ rate requires measuring normalized values to observe a stationary state.

\begin{figure}
\centering
 \includegraphics[width=\linewidth]{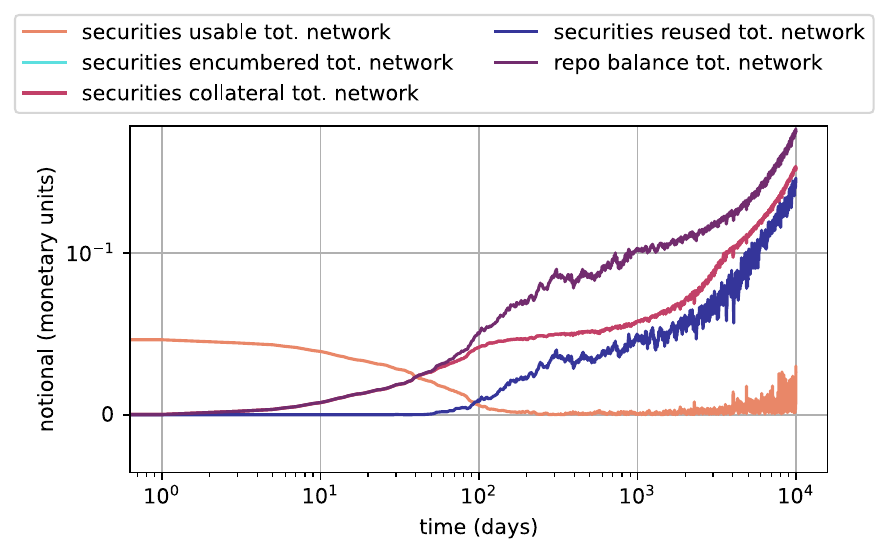}
 \caption{Time evolution of the collateral aggregates in the simulated banking system with $N=300$ banks, $g=0.04\%$, $v=5$, $\sigma=5\%$, $\beta_{\text{new}}=\beta=50\%$, and $\gamma_{\text{new}}=9\%$. Note that the securities encumbered are overlapping with the securities collateral.}
\label{fig:typical_run/collateral_aggregates}
\end{figure}

Figure~\ref{fig:typical_run/collateral_aggregates} displays a phase in which securities are consumed faster by the banking system than they are issued by the government, leading to a decrease of usable securities. When no securities remain, banks start to re-use collateral. The average rate of re-use converges to approximately $0.9$ (i.e., the typical length of a collateral chain is $2$, in line with the observations of \citet{LeCozEtAl-2024b}), because of the leverage constraint, which limits the balance sheet size of banks. 

\begin{figure} 
	\centering

 \subcaptionbox{Network density.\label{fig:typical_run/network_density}}{\includegraphics[width=\linewidth]{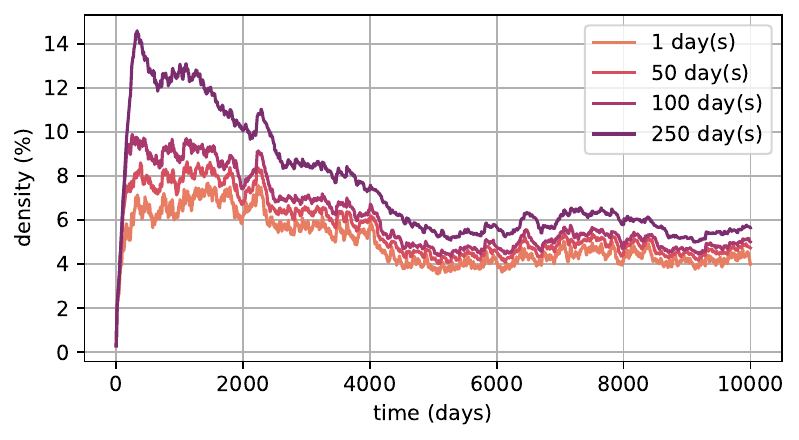}}
	
 \subcaptionbox{Jaccard network similarity index.\label{fig:typical_run/jaccard_index}}{\includegraphics[width=\linewidth]{ 
 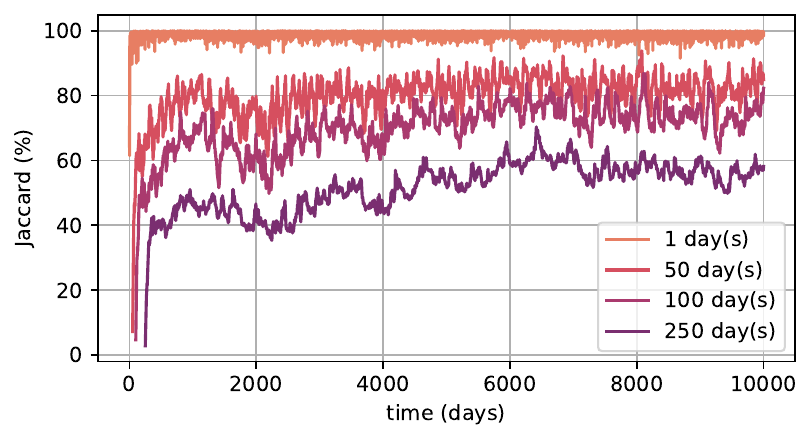}}
 
 \caption{Time evolution of the network density (\ref{fig:typical_run/network_density}) and the share of stable links from one period to another (\ref{fig:typical_run/jaccard_index}) in the simulated money markets with $N=300$ banks, $g=0.04\%$, $v=5$, $\sigma=5\%$, $\beta_{\text{new}}=\beta=50\%$, and $\gamma_{\text{new}}=9\%$. A link is defined as the existence of at least one repo over different aggregation periods, each corresponding to a given color. Pseudo-stationarity is reached after $6000$ steps due to the time of convergence toward a sufficiently unequal distribution of banks.}
\end{figure}

The model generates a relatively high density network compared to an unsecured lending network (see the ABM of \citet{Lux-2015}). Figure~\ref{fig:typical_run/network_density} shows a slow convergence of network density towards a regime close to stationarity because of the slow evolution of bank size distribution. Indeed, the heterogeneity in bank sizes never reaches a stationary state, even though the typical time scale required to observe non-stationarity is longer than our simulation window (see appendix~\ref{Random growth model}). We also find that some other combinations of input parameters lead the model with random growth (i.e., when $v \neq 0$) to generate a slowly increasing or decreasing density.

However, the model with equal growth rate (i.e. $v=0$) reaches a stationary state for a wider range of input parameters. In this case, the constantly increasing amount of new loans and deposits still slows down the convergence by increasing the average maturity of repos. However, a stationary regime is reached as long as the payment shocks are large enough to end these transactions (that is, $\sigma \ge 5\%$ according to our observations). Below this level, a stationary state can be reached by increasing the leverage constraint (i.e., a higher $\gamma^*$) or reducing the capital increase rate $\gamma_{\text{new}}$. This action would result in a reduction of collateral re-use (see section~\ref{Parameter space}).

We use the Jaccard network similarity index to characterize the stability of bilateral trading relationships from one period to another. The level of network stability exhibited in Fig.~\ref{fig:typical_run/jaccard_index} is consistent with observations on real financial networks \citep{Furfine-1999,AfonsoEtAl-2013,BlasquesEtAl-2018}. As mentioned above, a stationary state cannot be reached because of the slow evolution of bank size heterogeneity. Once again, this instability vanishes in the model with uniform growth rate (i.e. $v=0$).

\begin{figure}
\centering
 \includegraphics[width=\linewidth]{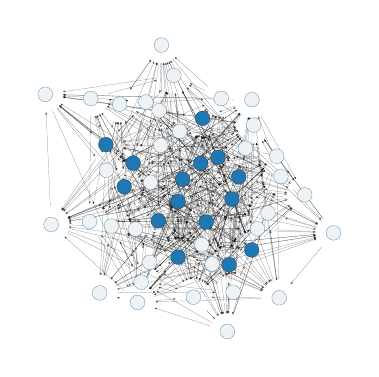}
 \caption{Core-periphery structure after $5000$ steps for $N=50$ banks, with $g=0.04\%$, $v=5$, $\sigma=5\%$, $\beta_{\text{new}}=\beta=50\%$, and $\gamma_{\text{new}}=9\%$. Links are defined through the aggregation of transactions that occurred within the last $50$ days. The core banks are identified through the method proposed by \citet{Lip-2011}, with a p-value of $10^{-10}$.}
\label{fig:typical_run/example_core_periphery}
\end{figure}

\begin{figure}
\centering
 \includegraphics[width=\linewidth]{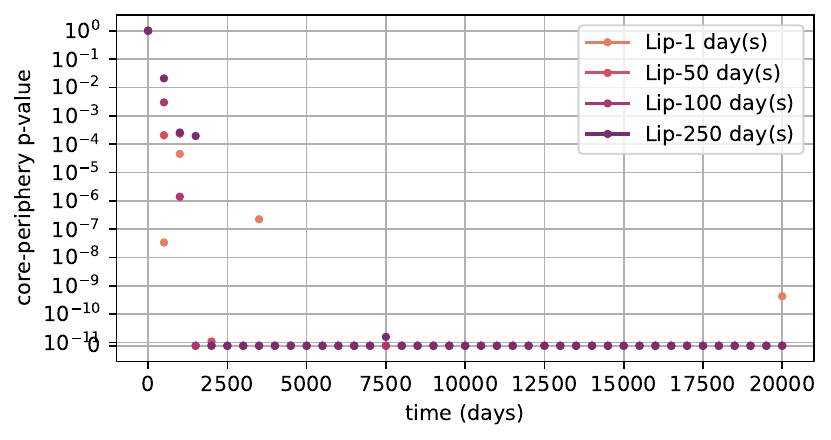}
 \caption{Time evolution of the p-values assessing the existence of a core-periphery structure according to the method proposed by \citet{Lip-2011} for $300$ banks, $g=0.04\%$, $v=5$, $\sigma=5\%$, $\beta_{\text{new}}=\beta=50\%$, and $\gamma_{\text{new}}=9\%$. A link is defined as the existence of at least a repo over different aggregation periods, corresponding to each color. The core-periphery structure emerges after $2500$ steps for all aggregation periods.}
\label{fig:typical_run/cpnet_pvalue-Lip}
\end{figure}

Figures~\ref{fig:typical_run/example_core_periphery} and~\ref{fig:typical_run/cpnet_pvalue-Lip} show that a core-periphery structure emerges from the network, even if the density is much higher than the one reported in~\citet{Lux-2015}. Notably, Fig.~\ref{fig:typical_run/cpnet_pvalue-Lip} reports the time evolution of the p-values from the Lip core-periphery test \citep{Lip-2011}: this kind of structure emerges after about $2500$ steps and is then stationary. However, other methods for assessing the significance of core-periphery \citep{BorgattiEverett-2000,BoydEtAl-2010,CucuringuEtAl-2016, RombachEtAl-2017, RossaEtAl-2013, KojakuMasuda-2018,KojakuMasuda-2018a}, based on different ways to characterize a core-periphery structure, do not lead to conclusive results. 

\begin{figure}
    \centering
 \includegraphics[width=0.75\linewidth]{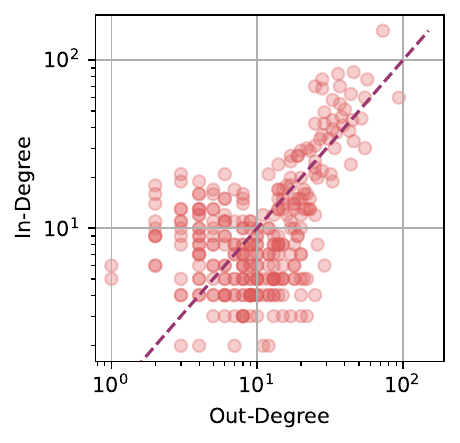}
 \caption{In--degree as a function of out--degree after $10 000$ steps, with $N=300$ banks, $g=0.04\%$, $v=5$, $\sigma=5\%$, $\beta_{\text{new}}=\beta=50\%$, and $\gamma_{\text{new}}=9\%$. Links are defined through the aggregation of transactions that occurred within the last $50$ days.}
\label{fig:typical_run/degree_distribution_on_day_9999}
\end{figure}

Finally, the generated network exhibits a slightly asymmetric in-- and out--degree distribution (see Fig.~\ref{fig:typical_run/degree_distribution_on_day_9999}), in line with the literature (see section~\ref{The interbank network topology}). 

\FloatBarrier
\subsection{Parameter space} \label{Parameter space}
Here, unless specified differently, we fix $g=0.04\%$, $v=0$ and $\nu=1.4$. It means that banks increase their balance sheet by $10\%$ a year ($1$ year $\approx 250$  steps) while keeping their size distribution (a power law with exponent $1.4$) constant. We also set $N=100$ and $\sigma=8\%$ to reach the stationary state faster. All other parameters are set as in the section~\ref{Typical run}. Each simulation is conducted over $10 000$ steps. For a given simulation, we define the stationary value of a given observable metric (for example, the network density) as its average across the last $200$ steps of the run. We simulate the same run (i.e. the same combination of input parameters) $100$ times. We finally report the mean over $100$ runs, excluding values outside of one standard deviation, of the stationary level of a given metric.

\paragraph{The effect of deposit outflow rate}
We assume $\beta=\beta_{\text{new}}$ in all simulations, ensuring that there is always enough collateral to meet the LCR needs of each bank.

\begin{figure} 
	\centering

 \subcaptionbox{Average regulatory ratios.\label{fig:parameter_space/accounting_view/regulatory_ratios/beta_reg}}{\includegraphics[width=0.7\linewidth]{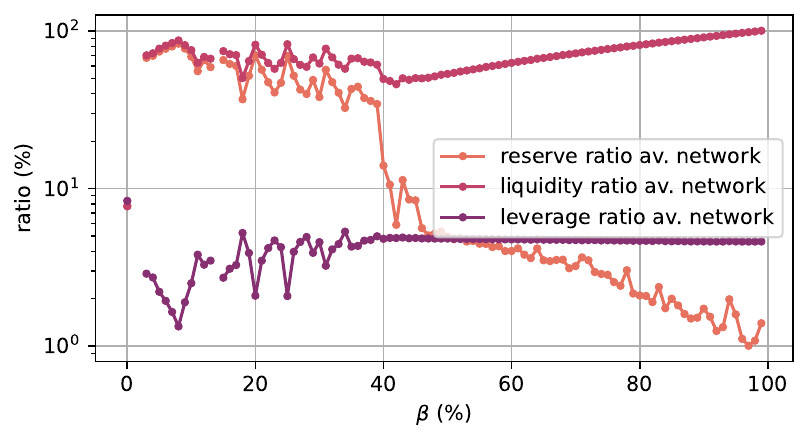}}
	
 \subcaptionbox{Collateral re-use.\label{fig:parameter_space/accounting_view/collateral_reuse/beta_reg}}{\includegraphics[width=0.7\linewidth]{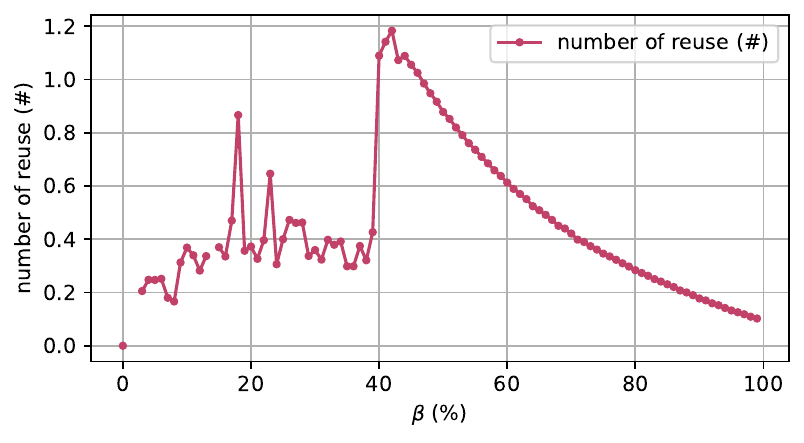}}
 
 \subcaptionbox{Network density.\label{fig:parameter_space/exposure_view/network_density/beta_reg}}{\includegraphics[width=0.7\linewidth]{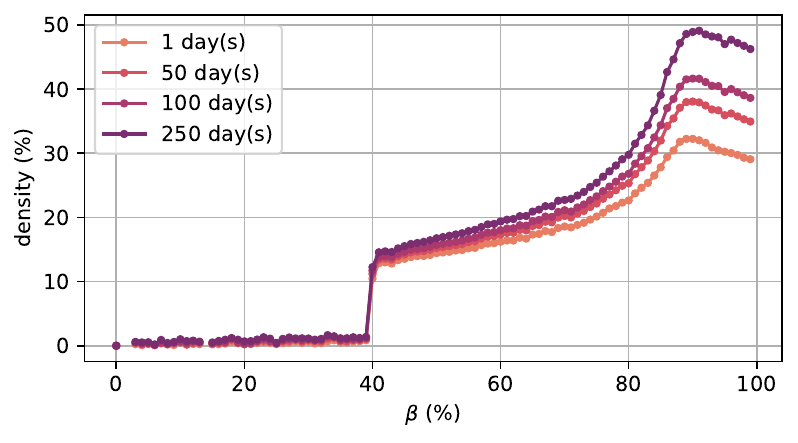}}
 
 \caption{Regulatory ratios, collateral re-use, and network density as a function of the deposit outflow rate~$\beta=\beta_{\text{new}}$, with $N=100$ banks, $g=0.04\%$, $\nu=1,4$, $\sigma=8\%$, and $\gamma_{\text{new}}=9\%$.}
\end{figure}

For high values of $\beta$ (i.e. $\geq 90\%$) there is much collateral available to absorb the payments shocks. This results in shorter collateral chains or lower re-use rate 
(Fig.~\ref{fig:parameter_space/accounting_view/collateral_reuse/beta_reg}). High values of $\beta$ are also associated to high network density (Fig~\ref{fig:parameter_space/exposure_view/network_density/beta_reg}) because the amount of repo required by each bank is proportional to $(1-\alpha) - (1-\beta)$ (see section~\ref{Banks' behavioral rules}). In other worlds, banks do not use central bank funding to manage their LCR ($\Delta M = (1-\beta) \ll 1$), therefore the excess liquidity is minimal (Fig.~\ref{fig:parameter_space/accounting_view/regulatory_ratios/beta_reg}) and even the smallest shocks must be absorbed on the repo market. 

When $\beta$ decreases ($\beta \in [40\%,90\%]$), banks have to rely more on central banking funding to manage their LCR, thus generating higher excess liquidity (Fig.~\ref{fig:parameter_space/accounting_view/regulatory_ratios/beta_reg}) and lower network density (Fig~\ref{fig:parameter_space/exposure_view/network_density/beta_reg}). One could have expected collateral re-use to decrease because excess liquidity reduces the effect of payment shocks. Yet, Fig.~\ref{fig:parameter_space/accounting_view/collateral_reuse/beta_reg} shows the opposite. This is because the total collateral available in the system starts becoming insufficient to cover all shocks, a state that we define as collateral scarcity. It does not mean that there is not enough collateral for all banks to meet their LCR requirements, but that the total amount of repo required to absorb payment shocks is higher than the available collateral. Banks react to collateral scarcity by increasing the length of collateral chains (Fig.~\ref{fig:parameter_space/accounting_view/collateral_reuse/beta_reg}) in line with empirical observations \citep{JankEtAl-2021}. We also observe a lower slope of the relationship between the density and the deposit outflow rate (Fig~\ref{fig:parameter_space/exposure_view/network_density/beta_reg}) for $\beta \in [50\%,75\%]$: this is because collateral scarcity reduces the chances of opening new repos, which allows existing repos to have a longer maturity.

However, for $\beta$ lower than $40\% \approx 5 \sigma $, there is not enough collateral in the banking system to absorb the payment shocks, so banks start relying on central funding for reserves management, generating high excess liquidity (Fig.~\ref{fig:parameter_space/accounting_view/regulatory_ratios/beta_reg}), low network density (Fig~\ref{fig:parameter_space/accounting_view/collateral_reuse/beta_reg}) and low collateral re-use (Fig.~\ref{fig:parameter_space/accounting_view/collateral_reuse/beta_reg}).

\begin{figure}
\centering
 \includegraphics[width=\linewidth]{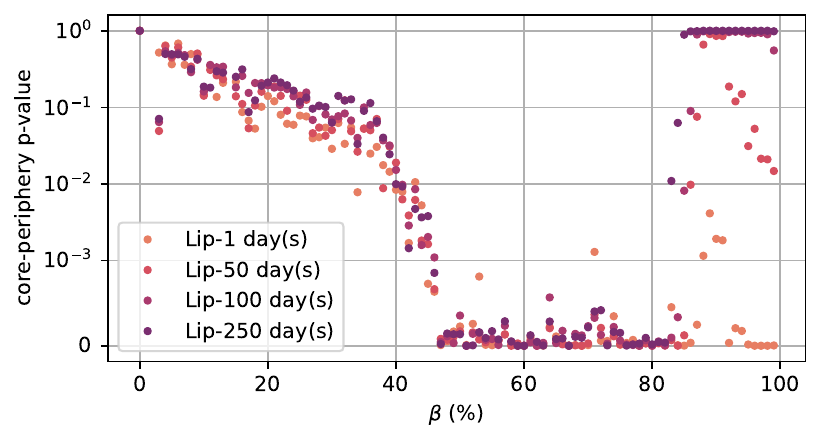}
 \caption{P-values assessing the existence of a core-periphery structure \citep{Lip-2011}, as a function of the deposit outflow rate~$\beta=\beta_{\text{new}}$, with $N=100$ banks, $g=0.04\%$, $\nu=1,4$, $\sigma=8\%$, and $\gamma_{\text{new}}=9\%$.}
\label{fig:parameter_space/exposure_view/core-periphery/beta_reg}
\end{figure}

Figure~\ref{fig:parameter_space/exposure_view/core-periphery/beta_reg} shows that the core-periphery structure is significant for $\beta$ in the range of $40$ to $80\%$. Outside of these limits, the density is either too high or too low to generate such a structure.

\paragraph{The effect of payment shocks' volatility}

\begin{figure} 
	\centering

 \subcaptionbox{Average maturity of repo transactions. \label{fig:parameter_space/transaction_view/repo_tr_mat_av_network/shocks_vol}}{\includegraphics[width=0.7\linewidth]{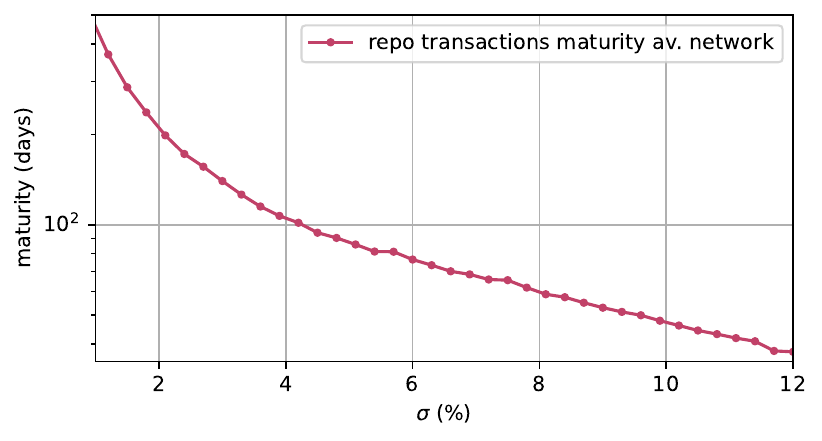}}
	
 \subcaptionbox{Network density.\label{fig:parameter_space/exposure_view/network_density/shocks_vol}}{\includegraphics[width=0.7\linewidth]{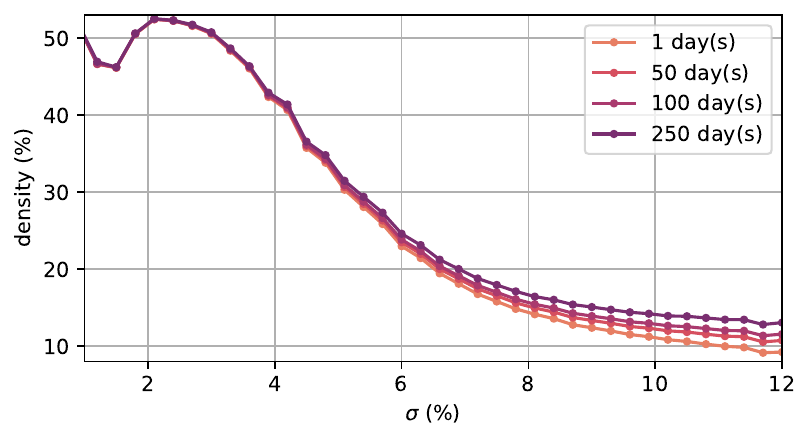 
 }}
 
 \subcaptionbox{Jaccard network similarity index.\label{fig:parameter_space/exposure_view/jaccard_index/shocks_vol}}{\includegraphics[width=0.7\linewidth]{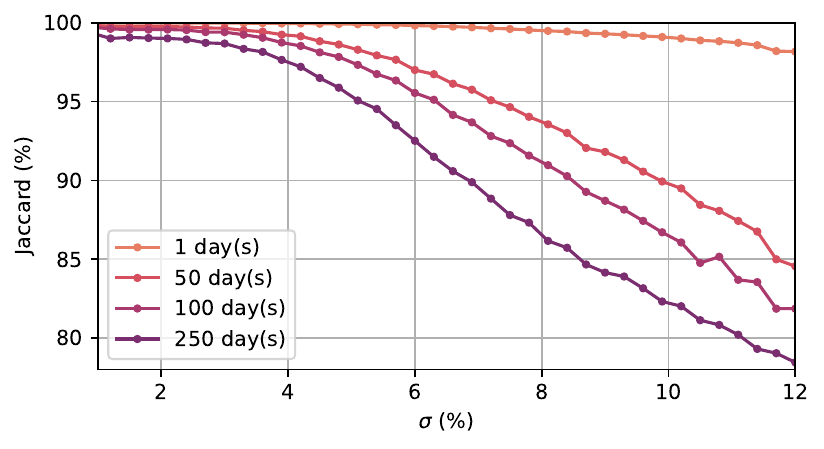}}
 
 \caption{Regulatory ratios, collateral re-use, and network density as a function of the volatility
of payment shocks $\sigma$, with $N=100$ banks, $g=0.04\%$, $\nu=1,4$, $\beta=\beta_{\text{new}}=50\%$ and $\gamma_{\text{new}}=9\%$. }
\end{figure}

\begin{figure}
\centering
 \includegraphics[width=\linewidth]{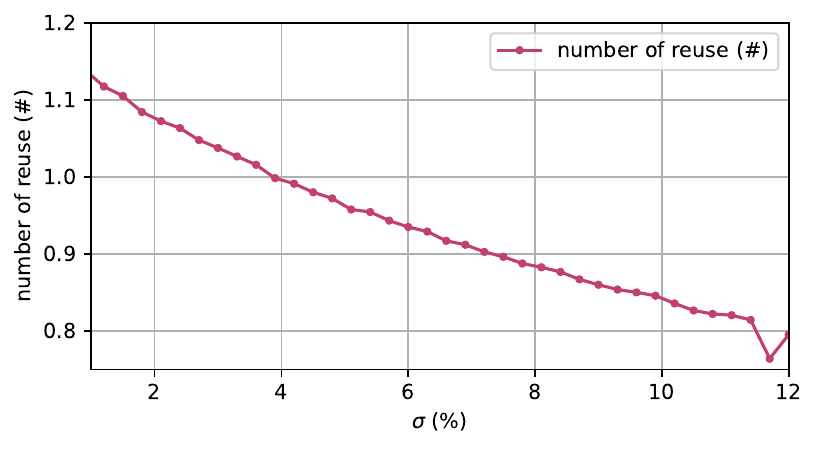}
 \caption{Collateral re-use as a function of the volatility of payment shocks~$\sigma$, with $N=100$ banks, $g=0.04\%$, $\nu=1,4$, $\beta=\beta_{\text{new}}=50\%$ and $\gamma_{\text{new}}=9\%$.}
\label{fig:parameter_space/accounting_view/collateral_reuse/shocks_vol}
\end{figure}

\begin{figure}
\centering
 \includegraphics[width=\linewidth]{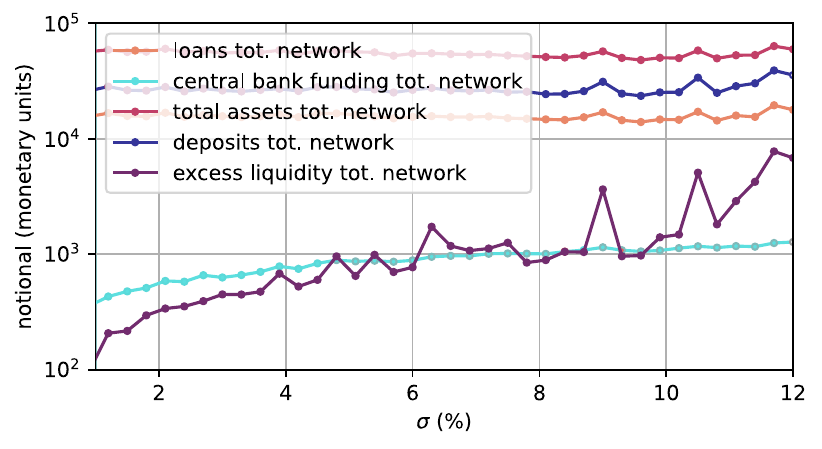}
 \caption{Macro-economic aggregates as a function of the volatility of payment shocks~$\sigma$, with $N=100$ banks, $g=0.04\%$, $\nu=1,4$, $\beta=\beta_{\text{new}}=50\%$ and $\gamma_{\text{new}}=9\%$.}
\label{fig:parameter_space/accounting_view/macro_economic_aggregates/shocks_vol}
\end{figure}

The lower the volatility of payment shocks, the higher the repo maturity as shown in Fig.~\ref{fig:parameter_space/transaction_view/repo_tr_mat_av_network/shocks_vol} (note the log-scale on the abscissa axis). In fact, a low volatility of deposits allows banks to hold their positions for longer periods. As a consequence, a low volatility of payment shocks is also associated with high network density (Fig.~\ref{fig:parameter_space/exposure_view/network_density/shocks_vol}), high Jaccard network similarity index (Fig.~\ref{fig:parameter_space/exposure_view/jaccard_index/shocks_vol}), and high collateral re-use rate (Fig.~\ref{fig:parameter_space/accounting_view/collateral_reuse/shocks_vol}). Conversely, the excess of liquidity in the banking system increases with the volatility of payment shocks (Fig.~\ref{fig:parameter_space/accounting_view/macro_economic_aggregates/shocks_vol}). Indeed, as explained in section~\ref{Banks' behavioral rules} banks' LCR management generates excess liquidity to absorb payment shocks.

\paragraph{Sensitivity analysis}
Appendix~\ref{Other parameter space analyses} shows the effect of three other control parameters: (i) the rate of capital increase, $\gamma_{\text{new}}$; the tail exponent, $\nu$, governing bank sizes heterogeneity; and the learning coefficient $\lambda$, controlling the speed at which trust levels are updated.

In particular, appendix~\ref{Other parameter space analyses} shows that the rate of collateral re-use is related to the ability of banks to increase their balance sheet size, which is tuned by $\gamma_{\text{new}}$.

\subsection{Stress testing} \label{Stress testing}
This model can be used to study the response of money markets to various stress scenarios. We name these scenarios after the most relevant crises recently faced by the banking system.  As in the previous section we fix $g=0.04\%$, $v=0$, $\nu=1.4$ and $N=100$. All the other parameters are fixed as in section~\ref{Typical run}.

\paragraph{Asset Purchase Program (APP)} This scenario corresponds to the disappearance of new collateral in the system, as it is bought by the central bank at issuance. Accordingly the parameter $\beta_{\text{new}}$ is set to $0$ between the steps $7000$ and $14000$. 

\begin{figure} 
	\centering

 \subcaptionbox{Time evolution of the main macroeconomic aggregates in the simulated banking system. \label{fig:APP/macro_economic_aggregates}}{\includegraphics[width=\linewidth]{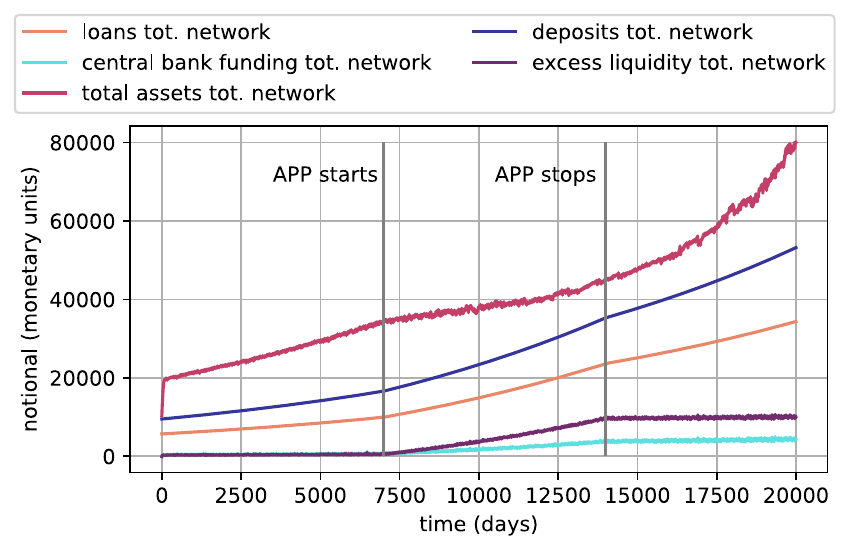}}
	
 \subcaptionbox{Time evolution of the collateral aggregates in the simulated banking system. Note that the securities encumbered are overlapping with the securities collateral. \label{fig:APP/collateral_aggregates}}{\includegraphics[width=\linewidth]{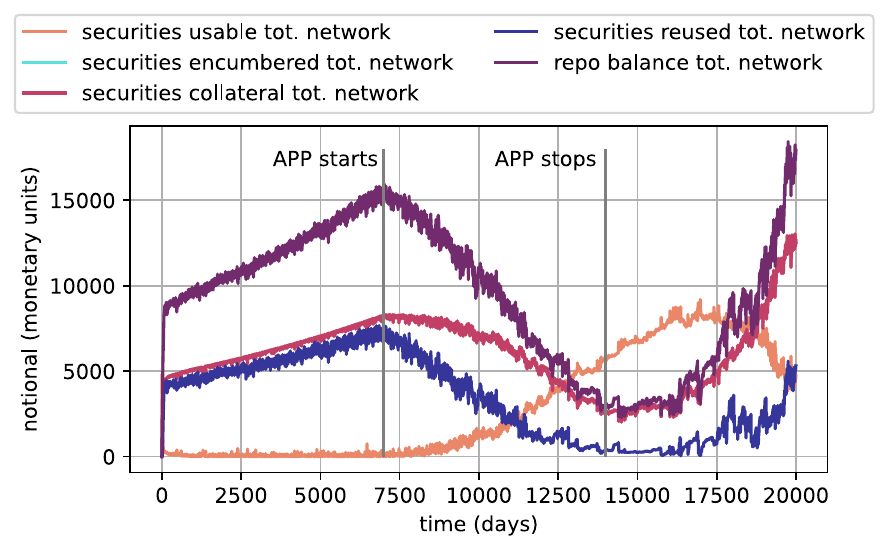 
 }}
 
 \caption{Asset Purchase Program. We assume the central bank buys all the newly created securities from the step $7000$ to $14000$, with $N=100$ banks, $g=0.04\%$, $\nu=1,4$, $\sigma=8\%$, and $\gamma_{\text{new}}=9\%$.\label{fig:APP_group_1}}
\end{figure}

\begin{figure} 
	\centering

 \subcaptionbox{Time evolution of the density in the simulated banking system.\label{fig:APP/network_density}}{\includegraphics[width=0.6\linewidth]{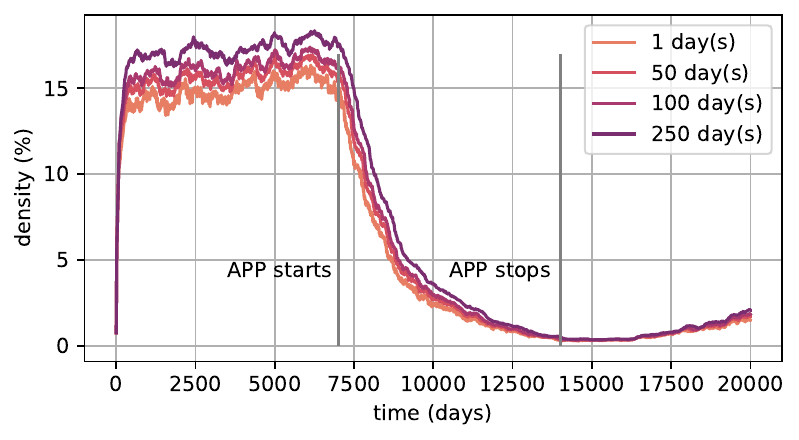}}
	
 \subcaptionbox{Time evolution of the number and notional of new repo transactions in the simulated banking system.\label{fig:APP/transactions}}{\includegraphics[width=0.6\linewidth]{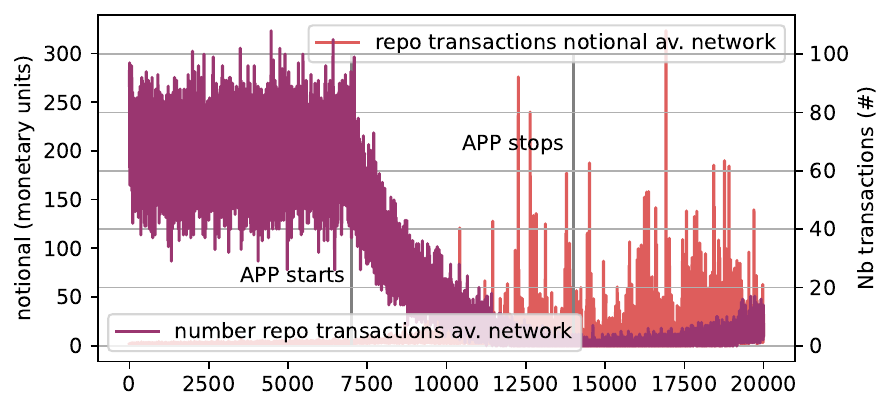}}

 \subcaptionbox{Time evolution of the p-values assessing the existence of a core-periphery structure according to the method proposed by \citet{Lip-2011}.\label{fig:APP/cpnet_pvalue-Lip}}{\includegraphics[width=0.6\linewidth]{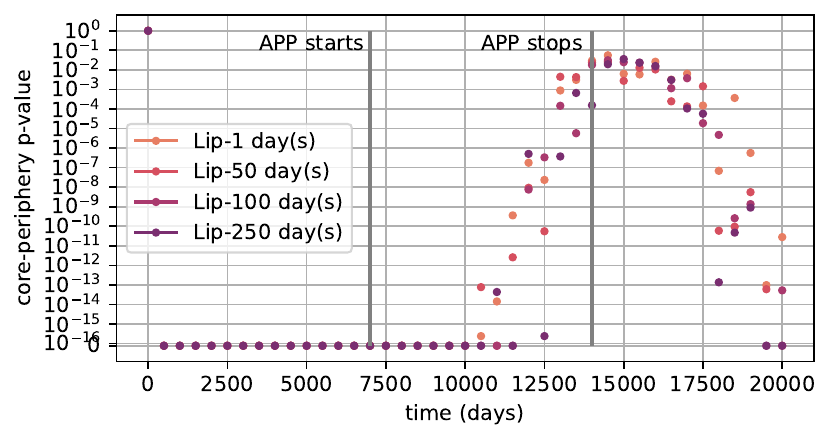}}
 
 \caption{Asset Purchase Program. We assume the central bank buys all the newly created securities from the step $7000$ to $14000$, with $N=100$ banks, $g=0.04\%$, $\nu=1,4$, $\sigma=8\%$, and $\gamma_{\text{new}}=9\%$. In Fig~\ref{fig:APP/network_density} and~\ref{fig:APP/cpnet_pvalue-Lip}, a link is defined as the existence of at least one repo over different aggregation periods, each corresponding to a given color. \label{fig:APP_group_2}}
\end{figure}

Figures~\ref{fig:APP_group_1} and \ref{fig:APP_group_2} show the impact of an APP on money markets. In essence, the APP provides money to the government that deposits this cash to the banking system, increasing excess liquidity (\ref{fig:APP/macro_economic_aggregates}). This excess of deposits reduces the need to access the interbank market, reducing the density of the network (\ref{fig:APP/network_density}) and the number of transactions (\ref{fig:APP/transactions}). Concurrently, some banks receiving large negative payment shocks need funding on the repo markets, increasing the average size of repo transaction (\ref{fig:APP/network_density}). Yet, some of them do not find sufficient collateral available due to the APP, hence must resort to central bank funding (\ref{fig:APP/macro_economic_aggregates}). Overall, the unwinding of existing repos for the bank receiving smaller shocks actually increases the amount of securities usable (\ref{fig:APP/collateral_aggregates}). The fall of money markets is almost complete at the end of the APP, which ultimately leads to the collapse of the core-periphery structure (\ref{fig:APP/cpnet_pvalue-Lip}). Note the long recovery of the market after step $14000$ (\ref{fig:APP/network_density}). Indeed, we did not simulate the maturing of existing collateral, which should mechanically decrease excess liquidity and reinforce the need for a repo market.

\paragraph{Great financial crisis.} This scenario corresponds to the default of a large bank due to the failure of all its loans to an economic agent. In such a case, a chain of collateral callbacks can be triggered by the counterparts of the defaulted bank. As all transactions are secured, there should be no contagion to the rest of the network. However, the economic agent records a loss equal to the amount of the defaulted loan due to the loss of its deposits and shares in the defaulted bank. The only consequence in our model is a loss of trust among all bank agents, leading banks to contact their counterparties randomly between steps 7000 and 14000. In other worlds, to model the great financial crisis in this framework, we do not have to model the default of a bank, but only its effect on the trust in the banking system, that is assumed to vanish for a given period before recovering.

\begin{figure} 
	\centering

 \subcaptionbox{Time evolution of the network density in the simulated banking system.\label{fig:GFC/network_density}}{\includegraphics[width=0.6\linewidth]{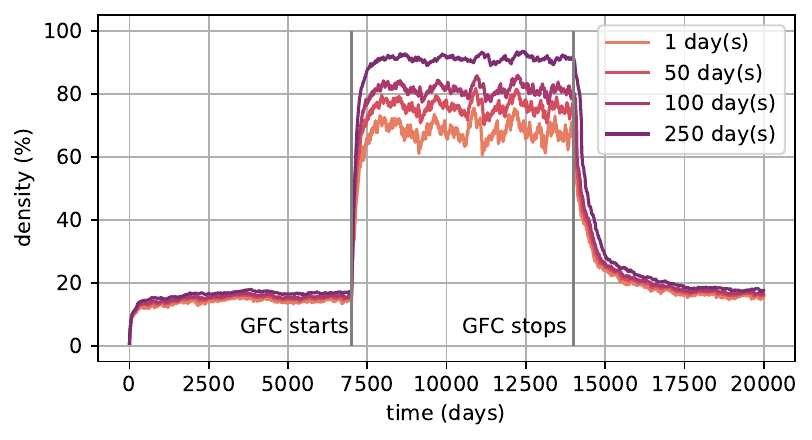}}
	
 \subcaptionbox{Time evolution of the Jaccard network similarity index in the simulated banking system.\label{fig:GFC/jaccard_index}}{\includegraphics[width=0.6\linewidth]{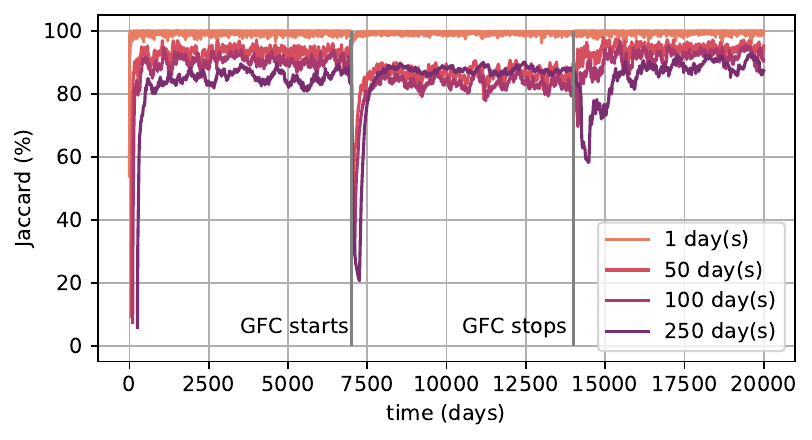}}

 \subcaptionbox{Time evolution of the number and notional of new repo transactions in the simulated banking system.\label{fig:GFC/transactions}}{\includegraphics[width=0.6\linewidth]{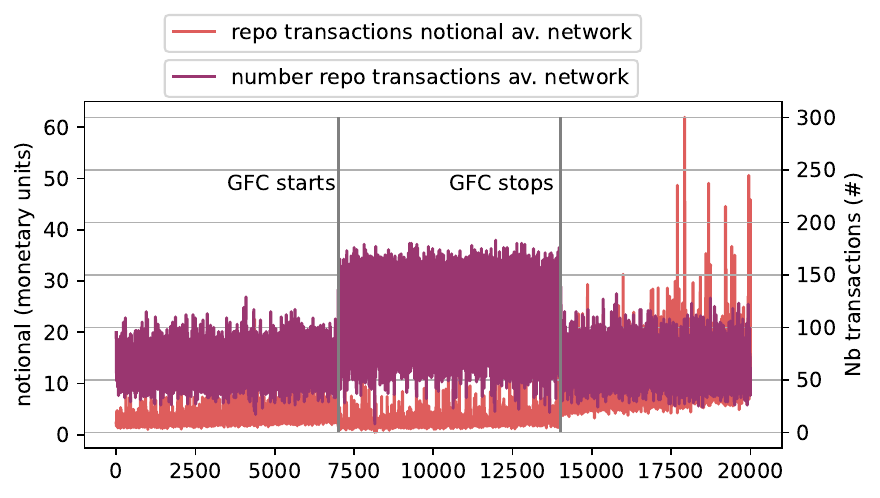}}
 
 \caption{Great financial crisis.We assume an absence of trust among banks from the step $7000$ to $14000$, with $N=100$ banks, $g=0.04\%$, $\nu=1,4$, $\sigma=8\%$, $\beta=\beta_{\text{new}}=50\%$ and $\gamma_{\text{new}}=9\%$. In Fig~\ref{fig:GFC/network_density} and~\ref{fig:GFC/jaccard_index}, a link is defined as the existence of at least one repo over different aggregation periods, each corresponding to a given color. \label{fig:GFC_group_1}}
\end{figure}

\begin{figure}
\centering
 \includegraphics[width=\linewidth]{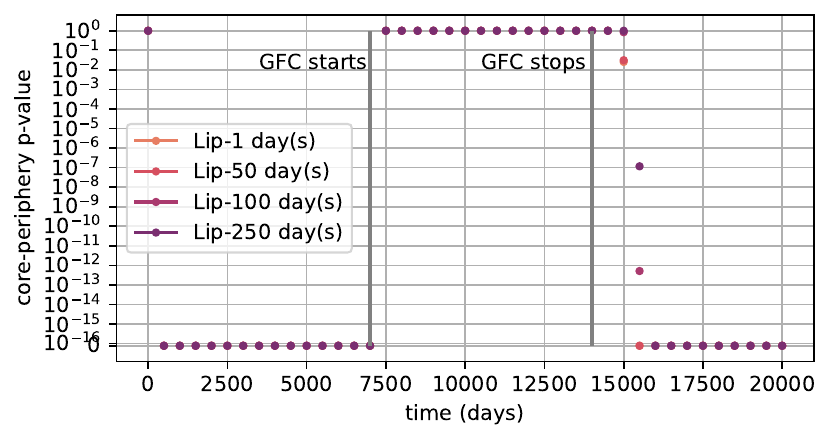}
 \caption{Great financial crisis. Time evolution of the p-values assessing the existence of a core-periphery structure according to the method proposed by \citet{Lip-2011}. We assume an absence of trust among banks from the step $7000$ to $14000$, with $N=100$ banks, $g=0.04\%$, $\nu=1,4$, $\sigma=8\%$, $\beta=\beta_{\text{new}}=50\%$ and $\gamma_{\text{new}}=9\%$. A link is defined as the existence of at least one repo over different aggregation periods, each corresponding to a given color.}
\label{fig:GFC/cpnet_pvalue-Lip}
\end{figure}

Fig.~\ref{fig:GFC_group_1} and \ref{fig:GFC/cpnet_pvalue-Lip} show the impact of such a scenario on money markets. Banks contact randomly their counterparties which increases the network density (\ref{fig:GFC/network_density}) and the number of transactions (\ref{fig:GFC/transactions}) but reduces the average notional of repo transactions (\ref{fig:GFC/transactions}). The network stability, measured by the Jaccard network similarity index, drops at the beginning of the crisis but quickly returns to almost its previous level (\ref{fig:GFC/jaccard_index}). The network is stable because banks are connected to almost all possible counterparties. As a consequence, the core-periphery structure vanishes (\ref{fig:GFC/cpnet_pvalue-Lip}). If we had added a minimum trust level for a transaction to occur, the market would have collapsed. There is no impact on macroeconomic and collateral aggregates.

\paragraph{Greek crisis} This scenario translates into a haircut on the collateral value. Thus, all cash lenders simultaneously request the posting of additional collateral. As not enough new collateral can be produced, borrowers of cash reimburse their existing repos and request more central bank funding. We leave for future work the development of a mechanism to account for daily margin calls and collateral value fluctuations in order to assess the consequences of such a scenario.

\paragraph{SVB bank run} This scenario materializes when a bank suddenly loses most of its deposits. In order to meet its regulatory constraint, such a bank would request large amounts of liquidity to the central bank and the other banks. In this model, this bank would survive the bank run because its liquidity need would be fulfilled by the access to infinite central bank funding. In practice, receiving central bank funding actually requires posting collateral, although it can be of lower quality than the one used in the repo markets. Hence, simulating such a crisis would require introducing a second type of collateral, which we leave for future work.

\section{Conclusion}

We have designed a minimal model of money market cash flows. In this approach, banks create money endogenously while absorbing payment shocks thanks to repo transactions. They respect reserves, liquidity, and leverage constraints. This framework sheds light on recent puzzles. Excess liquidity arises from the asymmetric responses of banks to payment shocks when managing their LCR. Banks cannot maintain both their reserves and LCR at their minimum levels and absorb daily payment shocks. Hence, excess liquidity should not disappear after the end of the APP (the sale of all its bonds by the ECB). Moreover, we find from our model that collateral is re-used due to the long canceling notice period of repos. Hence, reducing this practice would limit the ability of banks to manage short term liquidity needs. Collateral scarcity increases collateral re-use because positive shocks must be absorbed by more borrowers. However, below a certain amount of securities in the banking system, the repo market collapses. Stable bilateral trading relationships, asymmetric in-- and out--degree distributions, and a core-periphery structure emerge as the effect of trust among banks, similarly to the approach of \citet{Lux-2015} for unsecured markets.

We used this model to assess the impact of two stress scenarios: (i) the disappearance of new securities during an APP and (ii) the systematic loss of trust during the GFC. Our findings confirm a positive impact of the full allotment procedure and LCR regulation on the stability of money markets. We notably observe that, even if the repo market collapses, loan production is maintained. In addition, secured transactions ensure the absence of contagion of a defaulted bank.

This model is also a policy tool to simulate any changes in the allotment procedure of central banks or regulatory constraints' modifications. It shows that changing individual regulation can affect the system in an unintended way. Notably, setting low levels (around $5\sigma$) of the deposit outflow rate significantly increases excess liquidity and collateral re-use but reduces network density. If we decrease the amount of securities held by banks below the size of the largest payment shocks, i.e. $\beta \leq 3\sigma$, the repo market collapses and excess liquidity explodes.

 Interbank markets are more sensitive to liquidity risk than to interest rate risk because of the short maturity of exposures. However, introducing prices into this framework would allow one to model the transmission of central bank rates to money markets. Such a framework could explain another money market puzzle: the departure of the repo rates from the ECB's interest rate corridor \citep{PiquardSalakhova-2019}.

\section{Acknowledgments}
We are indebted to David Kass who contributed significantly to the building of the structure of the early Python code.  We extend our thanks to Thomas Lux and Stefano Corradin, who contributed to our research through fruitful discussions. Finally, we thank Bertrand Hassani and the ANRT (CIFRE number 2021/0902) for providing us with the opportunity to conduct this research at Quant AI Lab.

This research was conducted within the Econophysics \& Complex Systems Research Chair, under the aegis of the Fondation du Risque, the Fondation de l’École polytechnique, the École polytechnique and Capital Fund Management.
\clearpage

\FloatBarrier
\bibliographystyle{apsrev4-2} 
\bibliography{zotero}

\clearpage

\appendix
\section{Random growth model} \label{Random growth model}
\begin{figure}
    \centering
    \includegraphics[width=\linewidth]{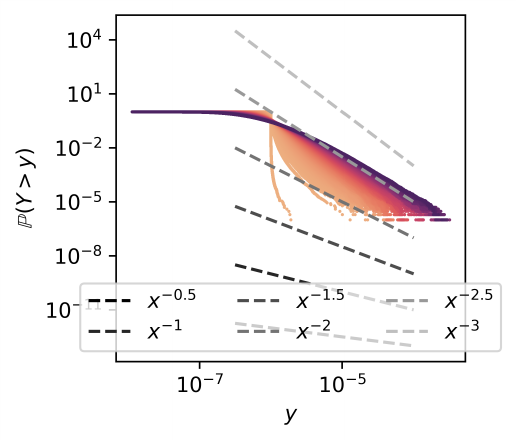}
    \caption{Evolution of cumulative distribution function of the relative sizes of banks across $100 \; 000$ days with an average growth rate $g=10\%$ and a volatility $v=10$. Each color corresponds to a snapshot every $500$ steps from the shortest (orange) to the latest (purple). Within a certain range, the measure of the tail exponent drops from infinity to around $3$ in $5000$ steps. This exponent is still $2.5$ after $100 \; 000$ steps.}
    \label{fig:results/99500_power_law}
\end{figure}

We want to model money creation through positive shocks fluctuating around an average rate $g$ of new money. Such model can be formulated by
\begin{align}
\label{eq:money_creation_Appendix}
    &X_i(t+1) = (g Z_i(t) +1) X_i(t), \nonumber \\
    &X_i(0) = x_0,
\end{align}
where $Z_i(t) = e^{\sigma_Z \epsilon_i(t) - \frac{1}{2}\sigma_Z^2}$. We recall $(\epsilon_i(t))$ are independent normalized centered Gaussian random variables across banks and time. Taking the expectation of Eq.~\eqref{eq:money_creation_Appendix} yields
\begin{align}
\label{eq:money_creation_mean}
    \langle X_i(t) \rangle = x_0 (1+g)^t.
\end{align}
Similarly, taking the expectation of the square of Eq.~\eqref{eq:money_creation_Appendix} gives
\begin{align}
\label{eq:money_creation_squared_mean}
    \langle X_i^2(t) \rangle = x_0^2 (g^2e^{\sigma_Z^2} + 2g +1)^t,
\end{align}
which shows $X(t)$ is non-stationary. For $t \gg 1$ and increments of small size $\Delta t$, by taking the logarithm of Eq.~\eqref{eq:money_creation_Appendix}, we have
\begin{align}
    \ln(X_i(t)) = \sum_{t'=0}^t\ln(gZ_i(t')+1) + \ln(x_0).
\end{align}
Assuming $g^2 e^{\sigma_Z^2} \ll 1$ (i.e. the mean growth is small compared to fluctuations), $\ln(gZ_i+1)$ can be approximated by $gZ_i$ which has a mean $g$ and a variance $g^2 (e^{\sigma_Z^2} -1)$ that we note $g^2 v^2$. Hence, for $t$ sufficient large, by the central limit theorem, the log-returns $\ln(X_i(t+\Delta t)) - \ln(X_i(t))$ behave as a Gaussian of mean $g \Delta t $ and variance $ g^2 v^2 \Delta t $. Thus, in the limit $\Delta t \ll 1$ and $t\gg 1$, the process $X_i(t)$ reads
\begin{align}
    X_i(t) = x_0 e^{gt -\frac{1}{2}g^2v^2 t +gvB(t)},
\end{align}
where $B(t)$ is a Brownian motion. One can check that this expression yields the mean and variance in Eq.~\eqref{eq:money_creation_mean} and \eqref{eq:money_creation_squared_mean} for $g\ll 1$. The limit distribution of this type of random processes has been studied among others in  \citet{MarsiliEtAl-1998,Gabaix-1999, Mitzenmacher-2004}. Unfortunately this process has no stationary limit unless we prevent the smallest banks to become smaller than a certain barrier. Indeed, as noticed by \citet{Mitzenmacher-2004}, the logarithm of the density distribution of $X_i(t)$, noted $f_{X(t)}(x)$ reads
 \begin{align}
     \ln(f_{X(t)}(x)) &= - \left(\frac{3}{2}-\frac{1}{gv^2}\right)\ln(x) - \frac{1}{2g^2v^2 t}\ln(x)^2 \nonumber\\
     &-\ln(\sqrt{2\pi}g^2v^2 t) - \frac{1}{2gv^2} + \frac{1}{4},
 \end{align}
which is clearly non stationary. We could hope solving this issue by defining a bounded variable $Y_i(t)$, the re-scaled money creation of the bank~$i$ by the sum of money creation of the other banks:
\begin{align}
    Y_i(t) = \frac{X_i(t)}{\sum_{i=0}^N X_i(t)}.
\end{align}
In the limit of large $N$, the sum $\sum_{i=0}^N X_i(t)$ can be approximated by its mean, as long as the variance of the sum is small compared to its mean. Using Eq.~\eqref{eq:money_creation_mean}, \eqref{eq:money_creation_squared_mean} and the central limit theorem for large $N$, this condition is meet if
\begin{align}
    \left(\frac{g^2e^{\sigma_Z^2} + 2g +1}{(1+g)^2}\right)^t \ll N.
\end{align}
In this limit, $\sum_{i=0}^N X_i(t) \approx N x_0 e^{gt}$, so the density distribution function $f_{Y(t)}$ of the normalized variables $Y_i$ reads
\begin{align}
     \ln(f_{Y(t)}(y)) &= - \frac{3}{2}\ln(y) - \frac{1}{2g^2v^2 t}\ln(y)^2 \nonumber \\
     &-\ln(\sqrt{2\pi}g^2v^2 t) + \frac{1}{4}.
 \end{align}
Thus, for large $t$ and large $y$ (precisely for $\ln(y) \ll \sqrt{t}$), the quadratic term becomes negligible so the variable $Y_i$ behaves similarly to a power law of exponent $0.5$. Yet, the term $-\ln(\sqrt{2\pi}g^2v^2 t)$ shows that most banks have a size becoming infinitely small. Correcting such behavior requires either defining a negative drift pushing bank sizes towards a barrier \citep{MarsiliEtAl-1998,Gabaix-1999} or to allow banks to exchange wealth \citep{BouchaudMezard-2000}. Both options are in contradiction with the requirements of our model. In practice, for the typical values of $g=10\%$ per year, and $v=10$ (i.e. some banks double their balance sheet in a year, while the median bank grows by $1\%$), we observe that the distribution of the $Y_i$ is almost stationary after $5000$ steps (i.e. $ \approx 20$ years if we count $250$ business days per year). Indeed, Fig.~\ref{fig:results/99500_power_law} shows the distribution function of the relative sizes of banks moves very slowly between $5000$ and $100 \; 000$ steps.

\section{Sensitivity analysis} \label{Other parameter space analyses}
As a complement to section~\ref{Parameter space}, we present here the influence of several other key control parameters. Unless otherwise specified, all parameters are set as in section~\ref{Parameter space}. Each simulation is also conducted over $10 000$ steps. As previously, we simulate the same run $100$ times and report the mean, excluding values outside of one standard deviation, of the stationary level of a given metric.

\begin{figure}
\centering
 \includegraphics[width=\linewidth]{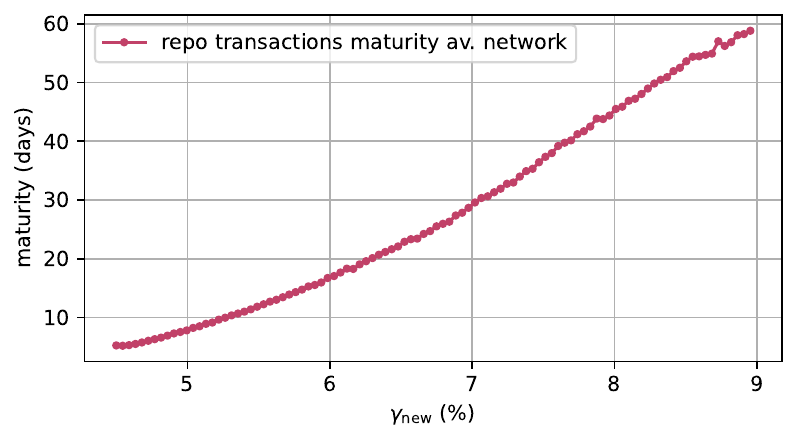}
 \caption{Average maturity of repo transactions as a function of the leverage ratio equivalent of new own funds~$\gamma_{\text{new}}$, with $N=100$ banks, $g=0.04\%$, $\nu=1,4$, $\sigma=8\%$, and $\beta=\beta_{\text{new}}=50\%$.}
\label{fig:parameter_space/transaction_view/repo_tr_mat_av_network/gamma_new}
\end{figure}

\begin{figure}
\centering
 \includegraphics[width=\linewidth]{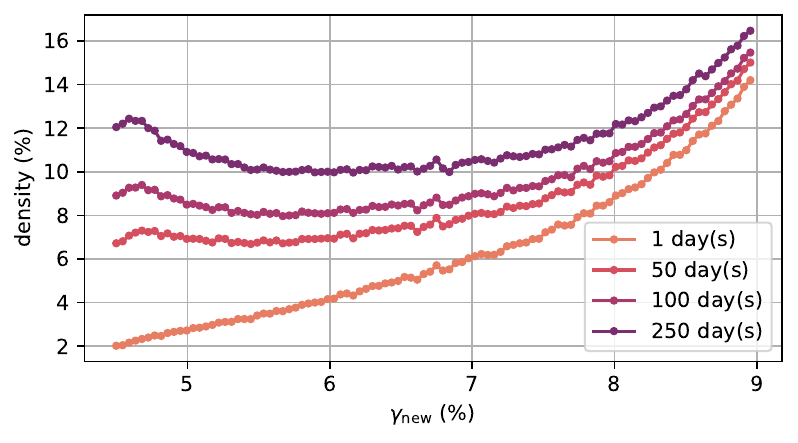}
 \caption{Network density as a function of the leverage ratio equivalent of new own funds~$\gamma_{\text{new}}$, with $N=100$ banks, $g=0.04\%$, $\nu=1,4$, $\sigma=8\%$, and $\beta=\beta_{\text{new}}=50\%$.}
\label{fig:parameter_space/exposure_view/network_density/gamma_new}
\end{figure}

\begin{figure}
\centering
 \includegraphics[width=\linewidth]{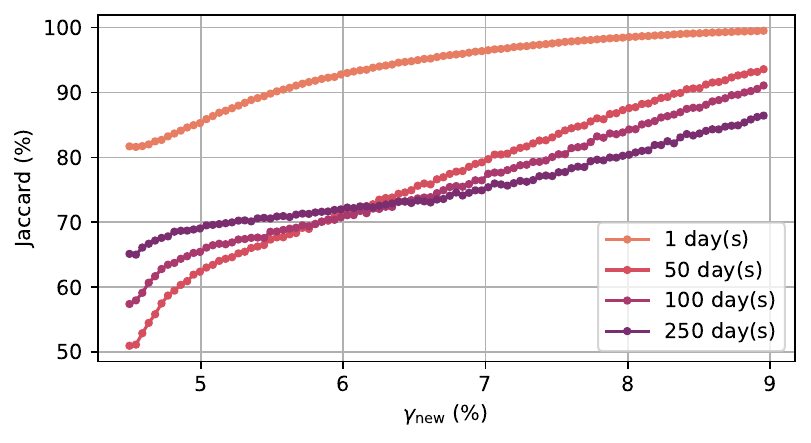}
 \caption{Jaccard network similarity index as a function of the leverage ratio equivalent of new own funds~$\gamma_{\text{new}}$, with $N=100$ banks, $g=0.04\%$, $\nu=1,4$, $\sigma=8\%$, and $\beta=\beta_{\text{new}}=50\%$.}
\label{fig:parameter_space/exposure_view/jaccard_index/gamma_new}
\end{figure}

\begin{figure}
\centering
 \includegraphics[width=\linewidth]{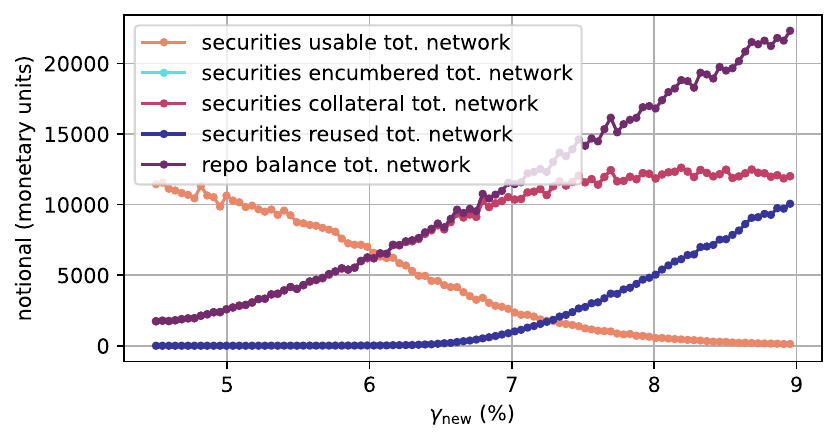}
 \caption{Collateral aggregates as a function of the leverage ratio equivalent of new own funds~$\gamma_{\text{new}}$, with $N=100$ banks, $g=0.04\%$, $\nu=1,4$, $\sigma=8\%$, and $\beta=\beta_{\text{new}}=50\%$. Note that the securities encumbered are overlapping with the securities collateral.}
\label{fig:parameter_space/accounting_view/collateral_aggregates/gamma_new}
\end{figure}

\paragraph{The effect of new own funds}
The leverage ratio is the constraint that limits the size of the balance sheet. Hence, the less binding the constraint (i.e. the higher the amount of new own funds $\gamma_{\text{new}}$ measured in leverage ratio equivalent, see section~\ref{Money creation shocks}), the longer the maturity of repos (Fig.~\ref{fig:parameter_space/transaction_view/repo_tr_mat_av_network/gamma_new}). This results in a higher network density (Fig.~\ref{fig:parameter_space/exposure_view/network_density/gamma_new}), a higher Jaccard network similarity index (Fig.~\ref{fig:parameter_space/exposure_view/jaccard_index/gamma_new}), and a higher rate of collateral re-use (Fig.~\ref{fig:parameter_space/accounting_view/collateral_aggregates/gamma_new}).

\begin{figure}
\centering
 \includegraphics[width=\linewidth]{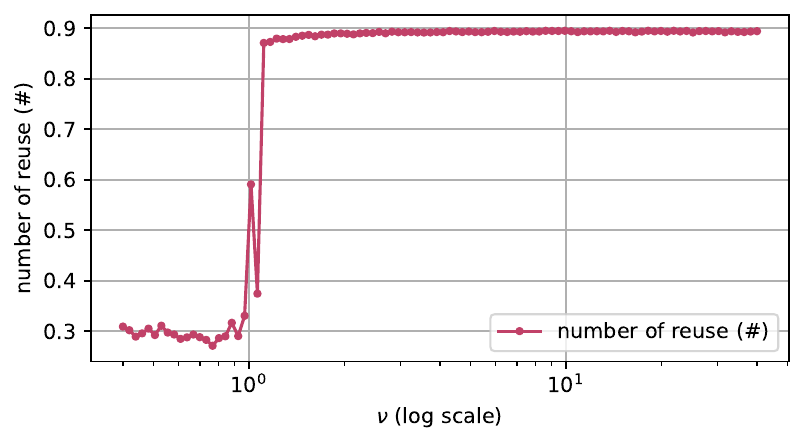}
 \caption{Collateral re-use as a function of the power law exponent~$\nu$ governing the distribution of bank sizes, with $N=100$ banks, $g=0.04\%$, $v=0$, $\sigma=8\%$, $\beta=\beta_{\text{new}}=50\%$, and $\gamma_{\text{new}}=9\%$.}
\label{fig:parameter_space/accounting_view/collateral_reuse/alpha_pareto}
\end{figure}

\begin{figure}
\centering
 \includegraphics[width=\linewidth]{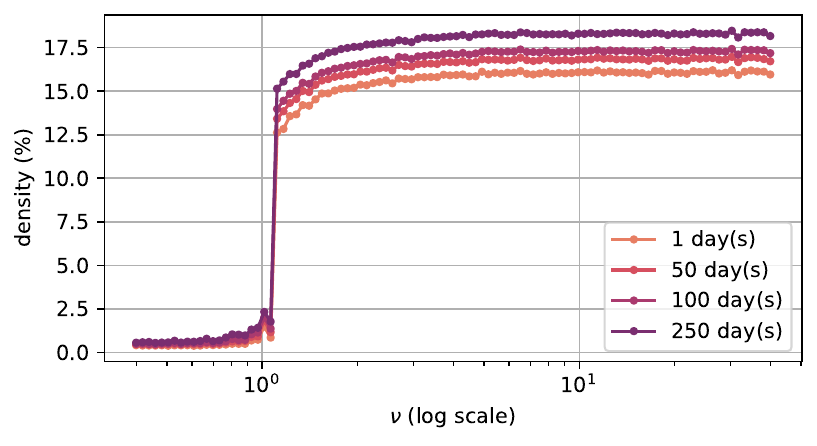}
 \caption{Network density as a function of the power law exponent~$\nu$ governing the distribution of bank sizes, with $N=100$ banks, $g=0.04\%$, $v=0$, $\sigma=8\%$, $\beta=\beta_{\text{new}}=50\%$, and $\gamma_{\text{new}}=9\%$.}
\label{fig:parameter_space/exposure_view/network_density/alpha_pareto}
\end{figure}

\begin{figure}
\centering
 \includegraphics[width=\linewidth]{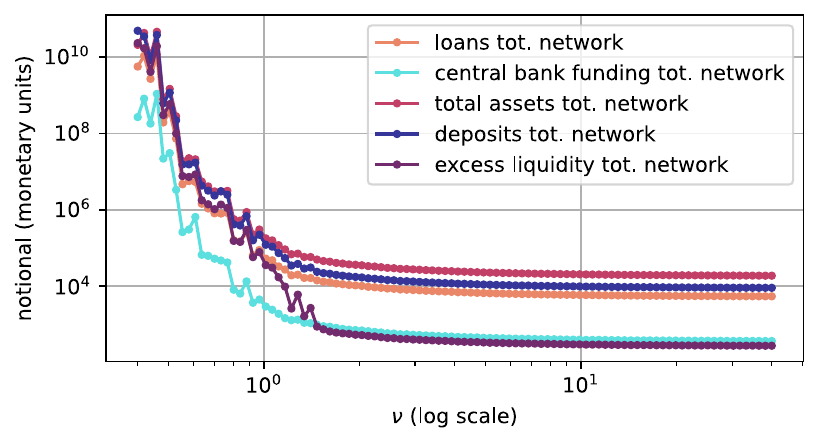}
 \caption{Macro-economic aggregates as a function of the power law exponent~$\nu$ governing the distribution of bank sizes, with $N=100$ banks, $g=0.04\%$, $v=0$, $\sigma=8\%$, $\beta=\beta_{\text{new}}=50\%$, and $\gamma_{\text{new}}=9\%$.}
\label{fig:parameter_space/accounting_view/macro_economic_aggregates/alpha_pareto}
\end{figure}

\paragraph{The effect of the size heterogeneity}
High levels of bank sizes' heterogeneity (i.e. low values of the tail exponent $\mu$) are associated to low collateral re-use rate (Fig.~\ref{fig:parameter_space/accounting_view/collateral_reuse/alpha_pareto}) and network density (Fig.~\ref{fig:parameter_space/exposure_view/network_density/alpha_pareto}). Indeed, when heterogeneity is high, the probability of a large positive shock to hit a large bank increases. This results in excess liquidity (Fig.~\ref{fig:parameter_space/accounting_view/macro_economic_aggregates/alpha_pareto}), which reduces the chances of subsequent shocks to generate liquidity needs, thereby reducing collateral re-use and network density.

\begin{figure}
\centering
 \includegraphics[width=\linewidth]{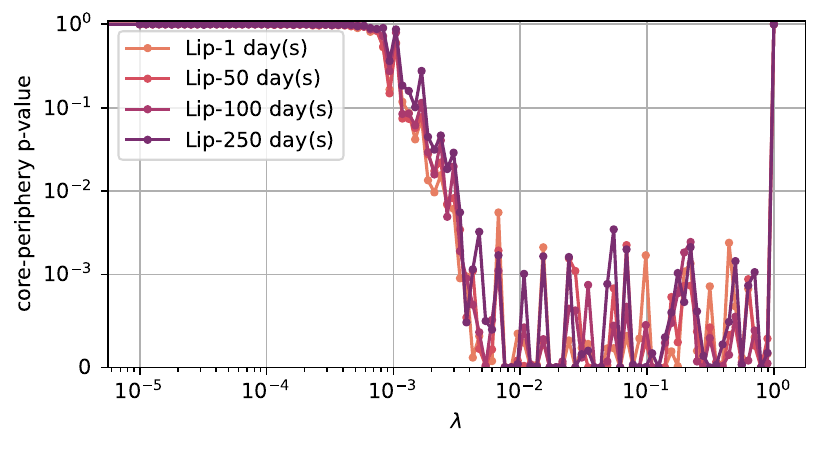}
 \caption{P-values assessing the existence of a core-periphery structure according to the method proposed by \citet{Lip-2011} as a function of the learning coefficient~$\lambda$, with $N=100$ banks, $g=0.04\%$, $\nu=1.4$, $\sigma=8\%$, $\beta=\beta_{\text{new}}=50\%$, and $\gamma_{\text{new}}=9\%$.}
\label{fig:parameter_space/exposure_view/core-periphery/learning_speed}
\end{figure}

\begin{figure}
\centering
 \includegraphics[width=\linewidth]{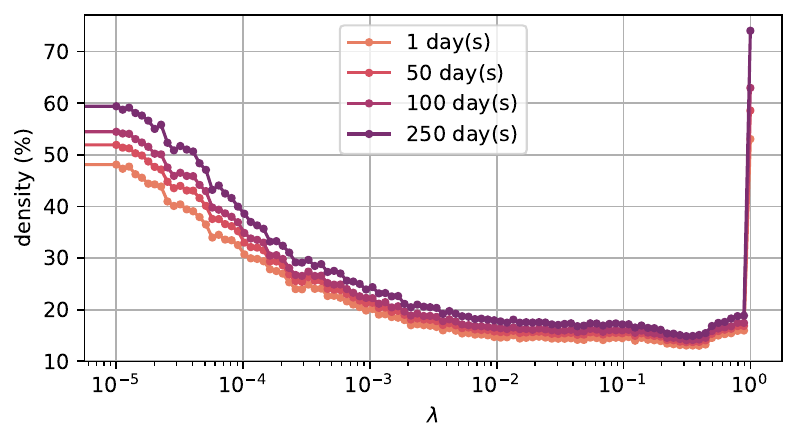}
 \caption{Network density as a function of the learning coefficient~$\lambda$, with $N=100$ banks, $g=0.04\%$, $\nu=1.4$, $\sigma=8\%$, $\beta=\beta_{\text{new}}=50\%$, and $\gamma_{\text{new}}=9\%$.}
\label{fig:parameter_space/exposure_view/network_density/learning_speed}
\end{figure}

\paragraph{The effect of the learning coefficient}
A core-periphery structure emerges if the learning coefficient~$\lambda$ is above a minimum level (around $0.01$ in Fig.~\ref{fig:parameter_space/exposure_view/core-periphery/learning_speed}). Below this value, banks do not learn enough quickly which counterparties to trade with, resulting in a high network density (Fig.~\ref{fig:parameter_space/exposure_view/network_density/learning_speed}).

\FloatBarrier

\end{document}